\documentclass[twocolumn]{aastex701}

\usepackage{amsmath}
\usepackage{graphicx}
\usepackage{url}
\usepackage{physics}
\usepackage{tikz}
\usepackage{tikz-3dplot}
\usetikzlibrary{backgrounds}

\newcommand{\NUPA}{\affiliation{ Department of Physics \& Astronomy, Northwestern University, Evanston, IL 60208, USA}}

\newcommand{\NUCIERA}{\affiliation{Center for Interdisciplinary Exploration \& Research in Astrophysics (CIERA), Northwestern University, Evanston, IL 60208, USA}}

\newcommand{\msun}{\mathrm{M}_{\odot}}

\newcommand{\tenc}{T_\mathrm{enc}}
\newcommand{\torb}{T_\mathrm{orb}}
\newcommand{\tgw}{T_\mathrm{GW}}
\newcommand{\tims}{T_\mathrm{IMS}}
\newcommand{\nims}{N_\mathrm{IMS}}
\newcommand{\tcross}{T_\mathrm{cr}}

\newcommand{\ave}[1]{\langle #1 \rangle}

\newcommand{\chieff}{\chi_\mathrm{eff}}
\newcommand{\chip}{\chi_\mathrm{p}}

\newcommand{\mbin}{m_\mathrm{b}}
\newcommand{\msin}{m_\mathrm{s}}
\newcommand{\mtot}{m_\mathrm{tot}}
\newcommand{\qbin}{q_\mathrm{b}}

\newcommand{\abin}{a_\mathrm{b}}
\newcommand{\ebin}{e_\mathrm{b}}
\newcommand{\lbin}{L_\mathrm{b}}
\newcommand{\lsin}{L_\mathrm{s}}
\newcommand{\ltot}{L_\mathrm{tot}}
\newcommand{\lbinvec}{\vec{L_\mathrm{b}}}
\newcommand{\lsinvec}{\vec{L_\mathrm{s}}}
\newcommand{\ltotvec}{\vec{L_\mathrm{tot}}}
\newcommand{\mubin}{\mu_\mathrm{b}}

\newcommand{\mutot}{\mu_\mathrm{tot}}
\newcommand{\Ebin}{E_\mathrm{b}}

\newcommand{\rp}{r_\mathrm{p}}
\newcommand{\rres}{r_\mathrm{res}}
\newcommand{\rexch}{r_\mathrm{exch}}
\newcommand{\vinf}{v_\mathrm{inf}}
\newcommand{\vcrit}{v_\mathrm{c}}
\newcommand{\niso}{n_\mathrm{iso}}
\newcommand{\nisojs}{n_\mathrm{iso,js}}
\newcommand{\nisoss}{n_\mathrm{iso,ss}}

\newcommand{\cosam}{\cos\alpha_m}
\newcommand{\mum}{\mu_m}

\newcommand{\vect}[1]{\boldsymbol{#1}}
\newcommand{\jhat}{\vect{\hat{j}}}
\newcommand{\shat}{\vect{\hat{s}}}

\newcommand{\rhat}{\vect{\hat{r}}}
\newcommand{\nhat}{\vect{\hat{n}}}
\newcommand{\xhat}{\vect{\hat{x}}}
\newcommand{\yhat}{\vect{\hat{y}}}
\newcommand{\zhat}{\vect{\hat{z}}}

\newcommand{\fewbody}{\texttt{Fewbody}\ }

\newcommand{\geodesicArc}[9]{%
  \draw[#9,line width=0.6pt]
    plot[domain=0:#6,smooth,samples=20,variable=\u]
    ({
      #5*(sin(#1)*cos(#2)*(1-\u)+sin(#3)*cos(#4)*\u)
      /sqrt(
        (sin(#1)*cos(#2)*(1-\u)+sin(#3)*cos(#4)*\u)^2+
        (sin(#1)*sin(#2)*(1-\u)+sin(#3)*sin(#4)*\u)^2+
        (cos(#1)*(1-\u)+cos(#3)*\u)^2)
    },{
      #5*(sin(#1)*sin(#2)*(1-\u)+sin(#3)*sin(#4)*\u)
      /sqrt(
        (sin(#1)*cos(#2)*(1-\u)+sin(#3)*cos(#4)*\u)^2+
        (sin(#1)*sin(#2)*(1-\u)+sin(#3)*sin(#4)*\u)^2+
        (cos(#1)*(1-\u)+cos(#3)*\u)^2)
    },{
      #5*(cos(#1)*(1-\u)+cos(#3)*\u)
      /sqrt(
        (sin(#1)*cos(#2)*(1-\u)+sin(#3)*cos(#4)*\u)^2+
        (sin(#1)*sin(#2)*(1-\u)+sin(#3)*sin(#4)*\u)^2+
        (cos(#1)*(1-\u)+cos(#3)*\u)^2)
    })
    node[#8] {#7}
    plot[domain=#6:1,smooth,samples=20,variable=\u]
    ({
      #5*(sin(#1)*cos(#2)*(1-\u)+sin(#3)*cos(#4)*\u)
      /sqrt(
        (sin(#1)*cos(#2)*(1-\u)+sin(#3)*cos(#4)*\u)^2+
        (sin(#1)*sin(#2)*(1-\u)+sin(#3)*sin(#4)*\u)^2+
        (cos(#1)*(1-\u)+cos(#3)*\u)^2)
    },{
      #5*(sin(#1)*sin(#2)*(1-\u)+sin(#3)*sin(#4)*\u)
      /sqrt(
        (sin(#1)*cos(#2)*(1-\u)+sin(#3)*cos(#4)*\u)^2+
        (sin(#1)*sin(#2)*(1-\u)+sin(#3)*sin(#4)*\u)^2+
        (cos(#1)*(1-\u)+cos(#3)*\u)^2)
    },{
      #5*(cos(#1)*(1-\u)+cos(#3)*\u)
      /sqrt(
        (sin(#1)*cos(#2)*(1-\u)+sin(#3)*cos(#4)*\u)^2+
        (sin(#1)*sin(#2)*(1-\u)+sin(#3)*sin(#4)*\u)^2+
        (cos(#1)*(1-\u)+cos(#3)*\u)^2)
    });
}

\begin{document}

\title{Getting Tilted: Random Walk of Binary Black Hole Spin-Orbit Alignment in Dense Star Clusters}
\shorttitle{BBH Spin-Orbit Random Walk}
\shortauthors{M.\ A.\ S.\ Martinez et al.}

\author[0000-0001-5285-4735]{Maia A. S. Martinez}
\NUPA
\NUCIERA
\email[show]{maiamartinez@u.northwestern.edu}

\author[0000-0003-3987-3776]{Christopher E. O'Connor}
\NUCIERA
\email{chrisoc@northwestern.edu}

\author[0000-0003-4412-2176]{Fulya Kıroğlu}
\NUCIERA
\email{fulya.kiroglu@northwestern.edu}

\author[0000-0002-7132-418X]{Frederic A. Rasio}
\NUPA
\NUCIERA
\email{rasio@northwestern.edu}

\correspondingauthor{Maia Martinez}

\begin{abstract}

It is commonly assumed that the spin-orbit angles of binary black holes (BBHs) originating from dense stellar environments rapidly converge to an isotropic distribution following a number of strong gravitational encounters. 
We challenge this assumption by modeling the evolution of the BBH orbital angular momentum orientation through successive binary--single encounters as a random walk on the unit sphere, yielding an exact solution for the orientation distribution after $n$ encounters and a closed-form expression for the number of encounters required to reach isotropy. 
To characterize the step distribution, we conduct a large suite of Newtonian point-particle scattering experiments with an equal-mass binary, varying the mass of the single, and obtain semi-analytic expressions for both the mean step size and the full distribution of steps. 
Applying these results to BBHs with initially aligned spins, as may arise from the evolution of primordial massive binaries, we find that spin-orbit alignment can survive several strong encounters before being erased. 
This has direct implications for the slight trend toward spin-orbit alignment reported in GWTC-5.0 as well as for the retention of hierarchical merger products, such as the components of GW231123.

\end{abstract}

\section{Introduction}

The recent release of GWTC-5.0 has brought the total number of observed gravitational-wave (GW) transients to $390$, the majority of them being binary black hole (BBH) mergers \citep{theligoscientificcollaborationGWTC50ObservationsSecond2026}.
Efforts to understand the formation of the merging BBHs include analyzing both the properties of individual exceptional events, such as GW231123 \citep{abacGW231123BinaryBlack2025}, and the statistical properties of the inferred astrophysical population \citep{theligoscientificcollaborationGWTC50PopulationProperties2026}.
One such property is the spin tilt ($\theta$) of BBH components, defined as the angle between the orbital angular momentum vector and the component spin vector.
The most recent analysis by the LIGO-Virgo-KAGRA (LVK) collaboration reports that at least $9\%$ of mergers must originate in environments with a preference for spin-orbit alignment, i.e. $\cos\theta>0$ \citep[see also][]{banagiriStructureSkewnessEffective2025}.
Moreover, \citet{theligoscientificcollaborationGWTC50PopulationProperties2026} reports that this finding is mass-dependent \citep[see also][]{rayAstrophysicalOriginBinary2026}.

This finding poses a challenge to the canonical formation scenarios of merging BBHs.
BBHs born from isolated binary stellar evolution are expected to have initially aligned spins  \citep{kalogeraSpinOrbitMisalignmentClose2000, farrDistinguishingSpinalignedIsotropic2017, gerosaSpinOrientationsMerging2018, wysockiExplainingLIGOsObservations2018, steinlePathwaysProducingBinary2021, callisterStateFieldBinary2021, stevensonBiasesEstimatesBlack2022}, though some workarounds have been suggested \citep{stegmannFlippingSpinsMass2021, taurisTossingBlackHole2022, baibhavRevisingSpinKick2024}.
The canonical dynamical channel, on the other hand, assembles BBHs through repeated gravitational encounters in dense stellar environments, naturally producing isotropic spin tilts \citep{rodriguez+2019, arcaseddaDRAGONIISimulationsIII2024}.
Alternative channels have been proposed in which gas-rich environments, such as the disks of active galactic nuclei (AGN), drive BBHs toward spin–orbit alignment \citep{mckernanLIGOVirgoCorrelationsMass2021,samsingAGNPotentialFactories2022, fabjSpinorbitMisalignmentsEccentric2026, tagawaPropertiesBlackHole2026}.

However, recent work has identified mechanisms by which BBHs in dense environments can develop a preference for spin-orbit alignment.
It is well known that massive stars are almost all born in binaries and higher-order multiples \citep{sanaBinaryInteractionDominates2012, moeMultiplicityStatisticsProperties2019, offnerOriginEvolutionMultiple2023}, and some fraction of BBH mergers in dense clusters may therefore come from primordial binaries rather than dynamically assembled ones. 
In fact, \citet{oconnorBlackHoleMergers2026} showed that in clusters with realistic initial binary fractions, some primordial BBHs, i.e., BBHs formed from a primordial binary, may indeed keep their original partners up until merger \citep[see similar findings by][]{hongBinaryBlackHole2018, arcaseddaDRAGONIISimulationsIII2024}.
Alternatively, when a dynamically assembled BBH collides with a star, the resulting accretion leads to aligned spins \citep{kirogluBlackHoleAccretion2025, kirogluHierarchicalMergersAccretiondriven2025, kirogluSpinOrbitAlignmentMerging2025}.
Regardless of the specific spin-alignment mechanism, BBHs in clusters harden through strong encounters with background singles, known as binary--single encounters, until ejected by dynamical recoil or until they merge \citep[e.g.,][]{heggieBinaryEvolutionStellar1975, antonini+rasio2016, samsing2018}.
\citet{oconnorBlackHoleMergers2026} show that, while some primordial BBHs may merge without any dynamical hardening and keep their original spin tilts, others may undergo several encounters before merging, changing the distribution of spin tilts.

Thus, we must understand the evolution of spin tilts due to successive binary--single encounters.
The few studies examining binary orientation changes due to encounters with comparable-mass singles have considered only individual encounters.
Namely, \citet{Stone+Leigh2019} used analytic methods to show that after a chaotic binary--single encounter, the binary orientation remains biased toward its original orientation.
Motivated by this observation, \citet{Trani+2021} conducted scattering experiments of binary--single encounters between BBHs and single BHs evolved with \texttt{BSE} and \texttt{SSE} \citep{hurley+2000, hurley+2002}.
They showed that primordial BBHs born via common envelope evolution in open clusters with small initial spin-orbit misalignment will typically undergo only one strong encounter before ejection from the host cluster.
However, in more massive clusters and for more general formation scenarios, BBHs may undergo several encounters before merging.
A number of studies have also examined the analogous problem of the rotational diffusion of massive BBH orientations due to encounters with test particles \citep{merrittRotationalBrownianMotion2002,gualandrisMassiveBlackHole2012,cuiOrbitalOrientationEvolution2014, rasskazovEvolutionBinarySupermassive2017, tomaselliSelfaccelerationHardeningBinaries2026}.
However, when the encountering singles are of comparable mass, the diffusion approximation breaks down, since a single encounter can produce large, discrete changes in the binary orientation.
Nevertheless, these studies motivate our approach: we model the cumulative evolution as an isotropic random walk on the unit sphere, the discrete analogue of rotational diffusion.

This problem also has relevance for the retention of BBH merger remnants.
If a BBH merges in a sufficiently massive cluster, the remnant may be retained and continue to pair up and merge with other BHs \citep[e.g., ][]{rodriguez+2019}.
The relative fraction of these subsequent mergers, known as hierarchical mergers, are one diagnostic by which the contribution of dynamical formation channels is constrained \citep[e.g.,][]{kimballEvidenceHierarchicalBlack2021,tongEvidencePairinstabilityGap2026,banagiriEvidenceThreeSubpopulations2025, rayReexaminingEvidencePairinstability2026, plunkettSignaturesSubpopulationHierarchical2026}.
However, spin-orbit alignment reduces the recoil kick magnitude, making retention of the merger remnant more likely. \citep{borchersGravitationalwaveKicksImpact2025, islamKickMattersImpact2026}.
If BBHs in clusters merge with preferentially aligned spins, then the expected fraction of hierarchical mergers will be higher than if the spin tilts are fully isotropic.
Furthermore, while it has been suggested that at least one component of GW231123 is a merger remnant \citep{2025ApJ...994L..54P, passengerGW231123HierarchicalMerger2026}, aligned spins are required to explain the large spin magnitudes of its components \citep{Stegmann2025}. 

In this work, we study the evolution of BBH spin-orbit misalignment due to successive binary--single encounters.
In Section~\ref{sec:theory}, we review relevant background information, present our random walk model, and present analytic results for the expected tilt from a single encounter.
In Section~\ref{sec:numerical_comparison}, we compare our analytic results to a large suite of scattering experiments and present relevant fits to our numerical results.
In Section~\ref{sec:discussion}, we demonstrate the relevance of our results for the population of merging LVK-type BBHs with an example calculation shown for GW231123-like systems and discuss future extensions of our work.
Finally, we summarize our results in Section~\ref{sec:summary}.

\section{Theoretical Background and Analytic Approximations}
\label{sec:theory}

\subsection{Astrophysical Preliminaries}
\label{sec:timescales}

We assume that the target BBH with total mass $\mbin=m_1 + m_2$ and mass ratio $\qbin=m_2/m_1\leq1$ has a semimajor axis $\abin$ and eccentricity $\ebin$, and it encounters a single background star with mass $\msin$ on a hyperbolic orbit with a relative velocity at infinity $\vinf$ and a distance of closest approach $\rp$. 
These relate to the impact parameter by the usual expression including gravitational focusing,
\begin{equation}
    b = r_{\rm p}\sqrt{1+\frac{2G\, \mtot}{r_{\rm p} v_{\inf}^2}}
    \label{eqn:b}
\end{equation}
where $\mtot$ is the total mass of the three body system.
The angular momenta of the binary and the single are $\lbinvec$ and $\lsinvec$ with magnitudes $\lbin = \mubin\sqrt{G\mbin\abin(1-\ebin^2)}$ and $\lsin = \mutot b \vinf$.
The energy of the binary is $\Ebin$.
It is typical to fix $\vinf$ as some fraction of
\begin{equation}
    \vcrit=\sqrt{\frac{G}{\mutot}\frac{m_1m_2}{\abin}},
    \label{eqn:vcrit}
\end{equation}
which is the relative velocity such that the total energy in the encounter is $0$.
For the remainder of this section, we restrict ourselves to the parabolic limit $\vinf \ll \vcrit$ since this applies to most star clusters of interest here.

Each individual encounter occurs on a timescale that can be characterized by the crossing time \citep[e.g.,][]{2006tbp..book.....V}
\begin{equation}
    \tcross \approx \frac{G\mtot^{5/2}}{2|\Ebin|^{3/2}} = \frac{1}{2\pi} \left( \frac{\mtot}{\sqrt{10m_1 m_2}} \right)^3 \torb,
    \label{eqn:tcross}
\end{equation}
where we approximated that the binary dominates the total energy of the encounter and $\torb$ is the orbital period of the original binary.
For encounters of comparable masses, $\tcross\lesssim\torb$.
While many interaction times are of order $\tcross$, some encounters are resonant and proceed on timescales much longer than $\tcross$.
During a resonant encounter, the single becomes temporarily bound to the binary and the three-body system alternates between a hierarchical state, where the system can be decomposed into an inner binary and an outer tertiary, and a ``democratic'' state, when there is no clean hierarchy.
\citet{samsingFormationEccentricCompact2014} have shown that the lifetime of these intermediate hierarchical states is
\begin{equation}
    \tims = \left( \frac{2}{1-1/a'} \right)^{3/2} \torb
\end{equation}
where $a'$ is the semimajor axis of the intermediate inner binary in units of $\abin$.
\citet{samsing2018} has shown that for the equal mass case, the average number of these intermediate states $\ave{\nims}\approx20$.
\citet{2025A&A...697A.118R} also showed that $\ave{\nims}$ is largest in the equal mass case, followed by the case when $m_2\approx\msin$.

Binaries in a cluster encounter single objects on an average timescale
\begin{equation}
    \tenc = (n\sigma\vinf)^{-1} \approx  \left(\frac{2\pi n G \mtot \rp}{\vinf} \right)^{-1}
    \label{eqn:tenc}
\end{equation}
where $n$ is the density of singles and we have used the parabolic approximation for the geometric cross section $\sigma=\pi b^2$.
These encounters will continue until the binary is either ejected due to a recoil kick from a hardening encounter or until the encounter time exceeds the GW merger timescale \citep{mandelAccurateAnalyticalFit2021}
\begin{multline}
    \tgw \approx \frac{5}{256}\frac{c^5}{G^3}\frac{\abin^4}{m_1 m_2 \mbin} \left(1-\ebin^2\right)^{7/2} \\
    \times \left( 1 + 0.27\ebin^{10} + 0.33\ebin^{20} + 0.2\ebin^{1000} \right) .
    \label{eqn:tgw}
\end{multline}
This expression fits the usual \citet{peters1964} result to within $3\%$ across the whole range of interest.

Finally, assume that the BHs have spins $\vect{s}_i$ with direction $\shat_i$ and magnitude $|\vect{s}_i|=\chi_i Gm_i^2/c$.
The spin tilts enter observationally through the effective spins $\chieff$ and $\chip$, defined as
\begin{equation}
    \chieff = \frac{m_1\chi_1\cos\theta_1 +m_2\chi_2\cos\theta_2}{m_1 + m_2},
\end{equation}
the mass-weighted sum of the spin components perpendicular to the orbital plane with $m_{1} \geq m_{2}$ by convention, and
\begin{equation}
    \chip = \max\left(\chi_1\sin\theta_1,\qbin\frac{4\qbin+3}{3+4\qbin}\chi_2\sin\theta_2\right)
\end{equation}
is the analogous quantity for the in-plane components with $q_{\rm b} = m_{2}/m_{1}$ the binary's mass ratio.
At lowest PN order, the spins will precess around $\jhat$ at an orbit-averaged frequency \citep[e.g.,][]{apostolatos+1994}
\begin{equation}
    \Omega_\mathrm{1(2)} = \frac{4\pi G\mubin}{c^2\abin(1-\ebin^2)}\frac{1}{\torb}\left(1 + \frac{3}{4}\frac{m_{2(1)}}{m_{1(2)}}\right).
    \label{eqn:prec_freq}
\end{equation}
For unequal mass binaries, the precession frequencies of the two spins differ by a relatively small factor.
The phase difference between the two spins can be characterized by the beat frequency
\begin{equation}
    \Omega_\mathrm{beat} = |\Omega_2-\Omega_1|=\Omega_1\left| 1 - \frac{1+\frac{3}{4\qbin}}{1 + \frac{3}{4}\qbin} \right|.
    \label{eqn:beat_freq}
\end{equation}
The effect of this precession is to introduce azimuthal phase mixing of the spin vectors.
Spin-spin coupling terms are also present, but we ignore them since $|\vect{s}_i|\ll\lbin$ during the dynamical hardening phase.
While these terms can be important in the final inspiral phase for determining the values of the final spin observables, the quantities $\chieff$ and $\chip$ are conserved during the inspiral at leading order \citep[e.g.,][]{gerosaMultitimescaleAnalysisPhase2015}.

\begin{figure}
    \centering
    \includegraphics[width=0.92\linewidth]{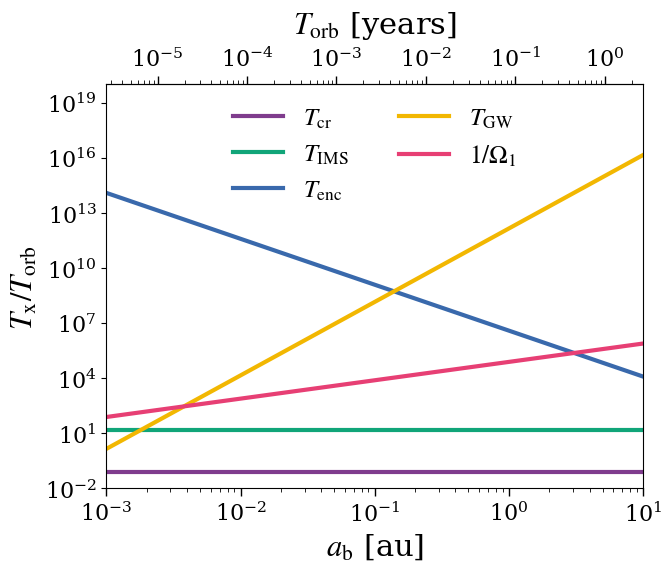}
    \caption{
    Comparison of timescales discussed in Sec.~\ref{sec:timescales} normalized by $\torb$ as a function of $\abin$ (bottom axis) and $\torb$ (top axis). 
    We made the following assumptions for the calculations.
    The binary properties were assumed to be $m_1=80\,\msun$, $m_2=60\,\msun$, and $\ebin=0.7$.
    For the cluster and encounter properties, we assumed $\vinf=30\,\mathrm{km/s}$, $n=10^6\,\mathrm{pc^{-3}}$, and $\msin=30\,\msun$.
    For the resonant hierarchical lifetime $\tims$, we assumed $a'=1.5$.
    For much of the dynamical hardening phase of a BBH, the spin phases are randomized by spin-orbit precession, but the spins may be treated as stationary during individual encounters.
    }
    \label{fig:timescales}
\end{figure}

In Figure~\ref{fig:timescales}, we compare some of these  timescales for a binary with $\mbin=140\,\msun$.
The assumed values are given in the figure caption; due to the large disparity of the timescales, the conclusions below are insensitive to the exact values chosen.
For most of a generic BBH's dynamical hardening phase, the relative phase of $\vect{s}_i$'s will be randomized between successive encounters.
However, during individual encounters, the $\vect{s}_i$'s can be treated as fixed.

\subsection{The Random Walk Model}
\label{sec:randomwalk}

\begin{figure}
    \centering
%
%

%
%

\tdplotsetmaincoords{70}{15}

\begin{tikzpicture}[tdplot_main_coords,scale=4,line cap=round,line join=round,every node/.style={font=\Large}]


\def\R{1}

\def\rsmall{0.32}

%
%
%

\def\tA{0}
\def\pA{0}

\def\tB{25}
\def\pB{-170}

\def\tC{35}
\def\pC{30}

\def\tD{80}
\def\pD{-45}

\def\tBa{35}
\def\pBa{-130}
\def\tBb{55}
\def\pBb{-160}
\def\tBc{75}
\def\pBc{-110}
\def\tBd{85}
\def\pBd{-60}



\foreach \t in {0,1,...,90} {

    \fill[gray!20,opacity=0.75]
        plot[domain=0:360,smooth,variable=\p]
        (
            {sin(\t)*cos(\p)},
            {sin(\t)*sin(\p)},
            {cos(\t)}
        );

}


\draw[dashed,gray,line width=0.6pt]
    plot[domain=\tdplotmainphi:\tdplotmainphi+180,smooth,variable=\p]
    ({cos(\p)},{sin(\p)},0);

\draw[line width=0.6pt, gray]
    plot[domain=\tdplotmainphi+180:\tdplotmainphi+360,smooth,variable=\p]
    ({cos(\p)},{sin(\p)},0);

\draw[gray,line width=0.6pt]
    plot[domain=-65:90,smooth,variable=\t]
    ({sin(\t)},0,{cos(\t)});

\draw[dashed,gray,line width=0.6pt]
    plot[domain=65:90,smooth,variable=\t]
    ({-sin(\t)},0,{cos(\t)});

\draw[dashed,gray,line width=0.6pt]
    plot[domain=90-\tdplotmaintheta:90,smooth,variable=\t]
    (0,{sin(\t)},{cos(\t)});

\draw[gray,line width=0.6pt]
    plot[domain=\tdplotmaintheta-90:90,smooth,variable=\t]
    (0,{-sin(\t)},{cos(\t)});


\draw[gray,dashed,line width=0.6pt]
    (-1,0,0) -- (1,0,0);

\draw[gray,dashed,line width=0.6pt]
    (0,-1,0) -- (0,1,0);


\coordinate (O) at (0,0,0);

\coordinate (A) at
({sin(\tA)*cos(\pA)},
 {sin(\tA)*sin(\pA)},
 {cos(\tA)});

\coordinate (B) at
({sin(\tB)*cos(\pB)},
 {sin(\tB)*sin(\pB)},
 {cos(\tB)});

\coordinate (C) at
({sin(\tC)*cos(\pC)},
 {sin(\tC)*sin(\pC)},
 {cos(\tC)});

\coordinate (D) at
({sin(\tD)*cos(\pD)},
 {sin(\tD)*sin(\pD)},
 {cos(\tD)});


\draw[->,line width=0.9pt] (O) -- (A)
node[above] {$\hat{j}_0$};

\draw[->,line width=0.9pt] (O) -- (B)
node[above] {$\hat{j}_1$};

\draw[->,line width=0.9pt] (O) -- (C)
node[above] {$\hat{s}_i$};

\draw[->,line width=0.9pt] (O) -- (D)
node[right] {$\hat{j}_n$};


\coordinate (As) at
({\rsmall*sin(\tA)*cos(\pA)},
 {\rsmall*sin(\tA)*sin(\pA)},
 {\rsmall*cos(\tA)});

\coordinate (Bs) at
({\rsmall*sin(\tB)*cos(\pB)},
 {\rsmall*sin(\tB)*sin(\pB)},
 {\rsmall*cos(\tB)});

\coordinate (Cs) at
({\rsmall*sin(\tC)*cos(\pC)},
 {\rsmall*sin(\tC)*sin(\pC)},
 {\rsmall*cos(\tC)});

\coordinate (Ds) at
({\rsmall*sin(\tD)*cos(\pD)},
 {\rsmall*sin(\tD)*sin(\pD)},
 {\rsmall*cos(\tD)});


\geodesicArc{\tA}{\pA}{\tB}{\pB}{0.4}{0.5}{$\alpha_m$}{above}{}
\geodesicArc{\tA}{\pA}{\tC}{\pC}{0.35}{0.75}{$\theta_{s,0}$}{above}{}
\geodesicArc{\tA}{\pA}{\tD}{\pD}{0.23}{0.7}{$\theta_{j,n}$}{right}{}
\geodesicArc{\tC}{\pC}{\tD}{\pD}{0.6}{0.55}{$\theta_{s,n}$}{right}{}

\geodesicArc{\tA}{\pA}{\tB}{\pB}{1}{0.5}{}{above}{dashed}
\geodesicArc{\tB}{\pB}{\tBa}{\pBa}{1}{0.5}{}{above}{dashed}
\geodesicArc{\tBa}{\pBa}{\tBb}{\pBb}{1}{0.5}{}{above}{dashed}
\geodesicArc{\tBb}{\pBb}{\tBc}{\pBc}{1}{0.5}{}{above}{dashed}
\geodesicArc{\tBc}{\pBc}{\tBd}{\pBd}{1}{0.5}{}{above}{dashed}
\geodesicArc{\tBd}{\pBd}{\tD}{\pD}{1}{0.5}{}{above}{dashed}

\end{tikzpicture}
    \caption{Diagram showing vectors and relative angles defined in Sec.~\ref{sec:randomwalk}. We assume that the angular momentum of the binary is initially pointing toward the North pole ($\jhat_0$). 
    The angular momentum orientation $\jhat_{1...n}$ takes discrete isotropic angular steps $\alpha_m$ due to encounters, which we depict with the dashed black line.
    Note that, aside from $\jhat_0$, the locations of the vectors are drawn from probability distributions. 
    For simplicity, we do not depict the time evolution of the spin vector $\shat_i$ since in general it depends on the time evolution of $\jhat_n$.}
    \label{fig:vector_diagram}
\end{figure}
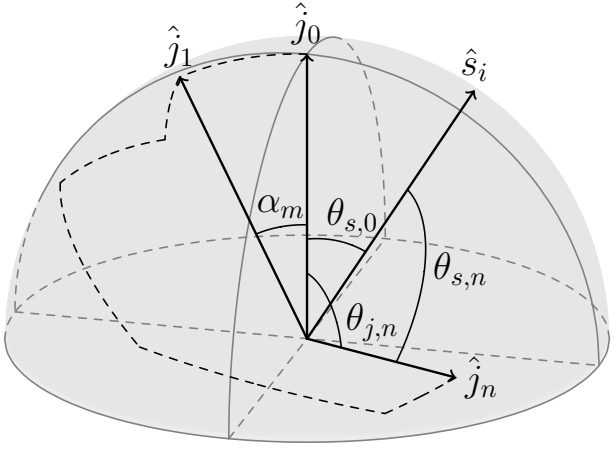

\subsubsection{Fixed Reference Vectors}

Figure~\ref{fig:vector_diagram} depicts the angles defined in the following discussion.
Let us assume that the initial orientation of the angular momentum $\jhat_0$ of $\mbin$ points to the North pole.
After a single encounter with $\msin$, $\vect{j}$ will be rotated in a random direction by some angle $\alpha_m$ drawn from a distribution $p(\mum) = p(\cosam)$.
Thus, the evolution of the orientation of $\jhat$ due to successive encounters of this kind is an isotropic random walk on the unit sphere \citep{1960RSPTA.252..317R}.
After some number of encounters $n$, the orientation of $\vect{j}$ with respect to its original orientation (i.e., the autocorrelation function) will be $\jhat_0\cdot\jhat_n = \cos\theta_{j,n} \equiv\mu_{j,n}$.
As shown by \citet{Kocsis+Tremaine2015}, the distribution of $\mu_{j,n}$ is 
\begin{equation}
    p(\mu_{j,n}) = \sum_{l=0}^{\infty} \frac{2l+1}{2} \ave{P_l(\mum)}^n P_l(\mu_{j,n}),
    \label{eqn:random_walk}
\end{equation}
where $P_l(\cdot)$ are the Legendre polynomials and $\ave{P_l(\mum)}$ are the Legendre moments calculated from the raw moments $\ave{\mum^n}$ of $p(\mum)$.

The limiting behavior of this distribution at late times is an isotropic distribution.
Each $P_l(\mu_n)$ mode has an associated characteristic decay scale $\tau_l = -{\ln\ave{P_l(\mu)}}^{-1}$.
For a reasonable distribution $p(\mum)$, the slowest evolving mode is the $l=1$ dipole mode, for which the associated Legendre moment is $\ave{P_1(\mum)}=\ave{\mum}$. 
We define the number of interactions required to roughly "isotropize" the orientation of $\jhat$ as
\begin{equation}
    \niso \approx 4\tau_1 = -\frac{4}{\ln\ave{\mum}}
    \label{eqn:decay_time}
\end{equation}
by computing $n$ such that the magnitude of the dipole term is less than $\sim2.7\%$ its initial value.
The decay timescale of the higher order terms of Eq.~\ref{eqn:random_walk} can be computed similarly and characterize the timescale of the erasure of small-scale angular structure in the distribution $p(\mu_n)$.
Note that in the limit of small angles, $-\ln{\ave{\mum}}\approx\ave{\alpha_m^2}/2$, the angular diffusion coefficient.

In Appendix~\ref{app:general_walk_reference}, we derive generalizations of Eq.~\ref{eqn:random_walk} and \ref{eqn:decay_time} for an arbitrarily distributed reference vector $\shat_{i}$:
\begin{equation}
    p(\shat_i\cdot\jhat_n) = \sum_{l=0}^\infty \frac{2l+1}{2} \zeta_l \ave{P_l(\mum)}^n P_l(\shat_i\cdot\jhat_n) 
    \label{eqn:random_walk_general}
\end{equation}
and
\begin{equation}
    \nisojs = -\frac{4 + \ln|\zeta_1|}{\ln \ave{\mum}},
    \label{eqn:decay_time_general}
\end{equation}
where $\zeta_l$ are the Legendre series expansion coefficients for the distribution of the reference vector, defined with respect to $\jhat_0$. 
In these expressions we've assumed that $p(\jhat_0\cdot\shat_{i})$ is azimuthally symmetric, though we show the more general form of the expression in App.~\ref{app:general_walk_reference}.
Since in general $|\zeta_l|\leq1$, the effect of using a reference angle other than $\jhat_0$ is to decrease the number of required encounters to isotropy, in accordance with expectations.

\subsubsection{Precessing Reference Vectors}

In our astrophysical scenario, the reference vectors $\shat_i$ are themselves not fixed as they are in the previous case.
This does not affect the validity of the previous expressions.
First, notice that instead of keeping $\shat_i$ fixed, one could equivalently keep $\jhat_i$ fixed and let $\shat_i$ evolve.
Now, returning to our original setup, assume for simplicity that $\jhat$ and $\shat_i$ are initially aligned.
After each step, $\shat_i$ will precess around $\jhat$.
Since $1/\Omega_i\ll\tenc$, we can assume that the phases are uniformly distributed.
Since this precession preserves the quantity $\jhat_n\cdot\shat_{i}$ in a manner that is azimuthally symmetric, Eqs.~\ref{eqn:random_walk_general} and \ref{eqn:decay_time_general} remain valid.

It is also of interest to know the evolution of the relative orientation of two different spin vectors evolving due to precession.
This is more difficult because the orientations of the two spin vectors are typically initially correlated.
The number of steps required to erase the initial correlation depends on the exact shape of the step distribution.
In principle, if one assumes that the number of steps required to erase the initial correlation is small, then the following results are obtained:
\begin{equation}
    p(\shat_{1,n}\cdot\shat_{2,n}) = \sum_{l=0}^\infty \frac{2l+1}{2} \zeta_l^2 \ave{P_l(\mum)}^{2n} P_l(\shat_{1,n}\cdot\shat_{2,n}) 
    \label{eqn:random_walk_ss}
\end{equation}
and
\begin{equation}
    \nisoss = -\frac{4 + 2\ln|\zeta_1|}{2\ln \ave{\mum}},
    \label{eqn:decay_time_ss}
\end{equation}
where we have introduced the subscript $n$ to make it clear that the positions of the spin vectors are evolving.
The derivation is almost identical to App.~\ref{app:general_walk_reference}.
This result is obtained by assuming that the new positions of both $\shat_i$ have Legendre series expansion coefficients identically given by Eq.~\ref{eqn:random_walk_general}, resulting in the power of $2$.
However, if the distribution of steps has significant weight towards small steps but a broad tail, then in some cases the correlation can persist for a long time relative to Eq.~\ref{eqn:decay_time_general} and Eq.~\ref{eqn:random_walk_ss} is incorrect.
In either case, the addition of precession causes the initial correlation of the spins to eventually isotropize.

\subsection{Tilts from a Single Encounter}

In order to fully characterize the evolution of the angular momentum vector due to successive encounters, it suffices to characterize the distribution $p(\mum)$ from a single encounter.
There are two regimes of interest.
When $r_p\gg\abin$, the binary is only slightly perturbed and the problem can be tackled with secular theory. 
However, when $r_p\approx\abin$, the binary is strongly perturbed and the interaction becomes chaotic.
We will examine these two regimes separately below.

\subsubsection{Weak Encounters}
\label{sec:weak_analytic}

In Appendix~\ref{app:quadrupole}, we show that at quadrupole order the change of the inclination due to a single encounter is
\begin{multline}
    \alpha_{m, \mathrm{w}} = \frac{15\pi}{16} \frac{\msin}{\mbin} \left( \frac{2\mbin}{\mtot}\right)^{1/2} \left(\frac{\abin}{\rp} \right)^{3/2} \\ \frac{\ebin^2}{(1-\ebin^2)^{1/2}} \sin{2i}\sin{2\omega}
    \label{eqn:delta_inc}
\end{multline}
where $i$ and $\omega$ are the standard Keplerian orbital elements defined with respect to $\lsinvec$.
Since these perturbations are small by construction, we approximate $\cos\alpha_{m, \mathrm{w}}\approx 1 - \alpha_{m, \mathrm{w}}^2/2$, giving
\begin{equation}
    \mu_{m, \mathrm{w}} = 1 - \left(\frac{15\pi}{16}\right)^2 \left(\frac{\abin}{\rp} \right)^3 \frac{\msin^2}{\mbin\mtot}
    \frac{\ebin^4}{1-\ebin^2}\sin^2{2i}\sin^2{2\omega}.
    \label{eqn:cos_delta_inc}
\end{equation}
Note that these expressions vanish when $\ebin=0$, for which the octupole order terms are necessary \citep{heggieEffectEncountersEccentricity1996}.
This expression also diverges in the limit $\ebin\rightarrow1$, for which second order perturbation techniques are necessary \citep{hamersAnalyticComputationSecular2019, hamersAnalyticComputationSecular2019a}.

For comparison, the mean change in energy is \citep{heggieBinaryEvolutionStellar1975}
\begin{equation}
    \ave{\Delta E_b} = -A\left(\frac{\rp}{\abin}\right)^{3/2}\exp\left[-B\left(\frac{\rp}{\abin}\right)^{3}\right],
    \label{eqn:energy}
\end{equation}
where
\begin{equation*}
    A = \xi2\sqrt{2}\frac{\mubin^{3/4}\mutot^{1/4}}{\mtot}\left(\frac{\vinf}{\vcrit}\right)^{3/2}|E_b|,
\end{equation*}
$\xi$ is an order-unity factor, and
\begin{equation*}
    B = \frac{1}{9} \frac{\mubin^{3/2} \mutot^{1/2}}{\msin^2} \left(\frac{\vinf}{\vcrit}\right)^{3}.
\end{equation*}
Due to the exponential term in Eq.~\ref{eqn:energy}, the change in energy falls more rapidly with increasing $\rp$ than the change in inclination. 
This is another manifestation of the old wisdom that it is easier to perturb the angular momentum of a binary than to perturb its energy.

For an isotropic distribution of encounters, $\omega$ and $\cos{i}$ are uniformly distributed between $[0,2\pi)$ and $[-1,1]$, respectively.
Because the two random variables are independent, we can express the raw moments $\ave{\mu_{m, \mathrm{w}}}$ in terms of the moments of $\ave{\sin^{2n}{2i}}$ and $\ave{\sin^{2n}{2\omega}}$.

Because $\omega$ is distributed uniformly, $\sin{2\omega}$ is distributed according to the Arcsine distribution bounded between $(-1,1)$. 
Similarly, $\sin^2{2\omega}\sim\mathrm{Arcsine}(0,1)$, which is a special case of the Beta distribution with $\alpha=\beta=1/2$.
From the definition of the Beta distribution, it can be shown that the moments are given by
\begin{equation*}
    \ave{\sin^{2n}{2\omega}} = \frac{B(1/2+n, 1/2)}{B(1/2, 1/2)} = \frac{(2n)!}{2^{2n}(n!)^2}
\end{equation*}
where $B(a,b)$ denotes the Beta function.
Similarly, $\sin^2{2i} = 4t(1-t)$, where $t=u^2\sim\mathrm{Beta(1/2,1)}$.
The moments can be computed similarly:
\begin{equation*}
    \ave{\sin^{2n}{2i}} = 4^n \frac{B(1/2 +n, 1+n)}{B(1/2,1)} = \frac{4^{2n}[(2n)!]^2}{(4n+1)!}.
\end{equation*}
As a result, we find that the raw moments of $\mum$ at fixed $\rp$ in the weak limit are
\begin{equation}
    \ave{\mum^n}_\mathrm{w} = \sum_{k=0}^{n}
    \binom{n}{k}
    (-C)^k \left(\frac{\abin}{\rp} \right)^{3k}.
    \label{eqn:weak_limit}
\end{equation}
with
\begin{equation}
    C = \frac{2^{2k}[(2k)!]^3}{(k!)^2 (4k+1)!}
    \left(\frac{15\pi}{16}\right)^2 \frac{\msin^2}{\mbin\mtot}
    \frac{\ebin^4}{1-\ebin^2}
    \label{eqn:C}
\end{equation}
For $n=1$, the factorial term evaluates to $4/15$.

One may also average this expression over a thermal eccentricity distribution. 
Due to the aforementioned unphysical divergence, the moments of the eccentricity term in $C$ are not finite.
However, eccentricities arbitrarily close to $1$ are unphysical in the sense that such binaries will quickly collide or circularize due to short range forces.
If we integrate the eccentricity term up to cutoffs in the range $0.99$ to $0.99999$, we find that the numerical value of the first moment is between $\sim3-10$, while the second and higher moments blow up rapidly.
This divergence is solely an artifact of first-order perturbation theory, as shown by \citet{hamersAnalyticComputationSecular2019, hamersAnalyticComputationSecular2019a}.
As we do not expect this small region of parameter space to significantly impact our numerical results, we leave a more rigorous extension including higher order terms to future work.

Equation~\ref{eqn:weak_limit} generalizes to a distribution of $b$ by integrating over $p(\rp)$.
In the parabolic limit, $b^2\propto\rp$ so that $p(\rp)$ is uniform over the range of interest.
With this, Eq.~\ref{eqn:weak_limit} can be "stitched together" with the results in the following section.

\subsubsection{Strong Encounters}
\label{sec:strong_analytic}

When $\rp$ is sufficiently small, the secular perturbative approximation fails.
In this regime, it is also possible for the encounter to become resonant.
During a resonant encounter, the binary undergoes chaotic dynamical evolution, well approximated as a random walk in $(\Ebin, \lbin)$, subject to the constraint that the total energy and angular momentum of the three-body system are preserved.
This random walk terminates once one of the objects in the encounter acquires enough energy to become unbound from the remaining binary.
This has been used to calculate statistical distributions of encounter outcomes \citep[e.g.,][]{Stone+Leigh2019, ginatAnalyticalStatisticalApproximate2021}.

It has also been established numerically that single escapers tend to escape with linear momentum concentrated near the plane perpendicular to the total angular momentum \citep{saslawGravitationalSlingshotStructure1974, valtonenStatisticsThreeBodyExperiments1974, anosovaDynamicalEvolutionEqualMass1986}.
The reason for this phenomenon comes from phase space arguments \citep{saariAngleEscapeThree1974, nashStatisticalTheoryDisruption1978}.
Roughly speaking, this constraint comes from the fact that while $\lsin$ is unbounded, $\lbin$ has a finite upper bound.
The upper bound corresponds to a zero-eccentricity orbit with an energy such that the single may be ejected. 
Thus, when there is a large total angular momentum, the single must carry away a large amount, requiring that the direction of escape is concentrated near the invariable plane \citep[see][for a more detailed pedagogical explanation]{2006tbp..book.....V}.

A related phenomenon was reported by \citet{Stone+Leigh2019}. 
By doing similar integrals over phase space, they found the following closed form expression which shows that for large $\ltot$, there is a strong bias towards prograde orbits:
\begin{equation}
    p(\mum) \propto \ln\left(\frac{L_C-\ltot\mum + \sqrt{L_C^2 + \ltot^2 -2\ltot\lbin\mum}}{\ltot(1-\mum)}\right).
    \label{eqn:stoneleigh}
\end{equation}
Here, $L_C = G\mubin\sqrt{m_1m_2\mbin/2|E_\mathrm{tot}|}$ and $E_\mathrm{tot}$ is the total energy of the three-body system.
While this expression is essentially exact and fits the numerical experiments conducted by \citet{Stone+Leigh2019} quite well, it is too unwieldy to use for the analytic calculations we require.

Motivated by this, we estimate the average inclination change due to the encounter at zeroth order.
Let us assume that the escaper must escape exactly within the invariable plane.
This requires that $\lbinvec$ and $\lsinvec$ are aligned at the end of the encounter.
Assume that initially, the angle between these two vectors is $\cos{i}$, which once again is distributed isotropically. 
Let 
\begin{equation}
    l = \frac{\lsin}{\lbin} = \frac{\msin}{\mubin}\sqrt{\frac{2\mbin}{\mtot}} \sqrt{\frac{\tilde{r}_p}{1-\ebin^2}}
    \label{eqn: l_ratio}
\end{equation}
where $\tilde{\rp}=\rp/\abin$.

The angle between $\lbinvec$ and $\ltotvec$ is
\begin{equation}
    \cos\alpha = \frac{1 + l\cos{i}}{\sqrt{1 + l^2 + 2l\cos{i}}}
\end{equation}
Evaluating this expression at $\cos{i}=\pm1$ gives the range of inclinations that give rise to escape.
Integrating over $\cos{i}$, we find
\begin{equation}
    \ave{\mum(l)}_\mathrm{s} = 
    \begin{cases}
        1 - \dfrac{1}{3}l^2, & 0\le l < 1, \\[6pt]
        \dfrac{2}{3l}, & l \ge 1.
    \end{cases}
    \label{eqn:mu_fixed_l}
\end{equation}
This behavior matches intuition.
In the limit of low $l$, $\lbinvec$ dominates the total angular momentum so the angle $\mum$ will tend to $1$, down to $l=0$ where $\ltotvec=\lbinvec$. 
As $l$ increases, the $\lsinvec$ dominates the total angular momentum, so the average converges to that of a purely isotropic distribution.

The above holds for a fixed angular momentum ratio, which depends on  $\rp/\abin$ and $e$. 
For our case, we must average over a suitable distribution of $l$. 
First, consider the case of fixed eccentricity.
Since $b^2\propto\rp$, the quantity $\sqrt{\rp}$ is distributed thermally for a distribution of encounters.
Let $\tilde{l}=l(r_{p,\mathrm{max}})$ and parameterize $l=\tilde{l}t$, where $t\sim\mathrm{Beta(2,1)}$ (i.e., thermally distributed). 
We integrate Eq.~\ref{eqn:mu_fixed_l} and find
\begin{equation}
    \ave{\mum(\tilde{l})}_\mathrm{s,\,fixed\,e} = 
    \begin{cases}
        1 - \dfrac{1}{6}\tilde{l}^2, & 0\le \tilde{l} < 1, \\[6pt]
        \dfrac{4}{3\tilde{l}} - \dfrac{1}{2\tilde{l}^2}, & \tilde{l} \ge 1.
    \end{cases}
    \label{eqn:mu_fixed_e}
\end{equation}

Second, consider a thermal eccentricity distribution.
In this case, both $\sqrt{\rp}$ and $\sqrt{1-\ebin^2}$ are thermally distributed.
Similar to before, let $\tilde{l}=l(r_{p,\mathrm{max}}, \ebin=0)$ and parameterize $l=\tilde{l}t$, where $t\sim\mathrm{Ratio(X,Y)}$ and $X, Y\sim\mathrm{Beta(2,1)}$.
In this case, we find
\begin{equation}
    \ave{\mum(\tilde{l})}_\mathrm{s,\,thermal} = 
    \begin{cases}
        1 + \tilde{l}^2\left(\dfrac{1}{3}\ln\tilde{l} - \dfrac{13}{36}\right), & 0\le \tilde{l} < 1, \\[6pt]
        \dfrac{8}{9\tilde{l}} - \dfrac{1}{4\tilde{l}^2}, & \tilde{l} \ge 1.
    \end{cases}
    \label{eqn:mu_thermal}
\end{equation}

We now comment on the validity of the previous calculation.
We began by assuming that following the encounter, both $\lsinvec$ and $\lbinvec$ become perfectly aligned with $\ltotvec$.
While time-reversal symmetry forbids this without dissipation, the approximation is reasonable as long as escape angles are sufficiently concentrated near the invariable plane.
However, since the actual distribution becomes less concentrated for smaller $\ltot$ normalized to its maximum value, we expect that the higher moments calculated with this approximation will be incorrect and do not attempt to calculate them here.
Moreover, the ergodic assumption underlying both Eq.~\ref{eqn:stoneleigh} and our calculation does not always hold.
As pointed out by \citet{Stone+Leigh2019} and studied in detail later by \citet{traniIslesRegularitySea2024a}, a large fraction of all strong encounters are direct, i.e. not chaotic.
Nevertheless, as we will show in the next section, our rough estimate captures the mass dependence of the numerical experiments surprisingly well.

\section{Comparison with Numerical Experiments}
\label{sec:numerical_comparison}

We conduct ensembles of binary--single scattering experiments with the numerical toolkit \fewbody \citep{fregeauStellarCollisionsBinarybinary2004}.
\fewbody integrates the Newtonian equations of motion with the eighth-order Runge-Kutta Dormand-Prince integration method with ninth-order error estimation and adaptive time steps.
We set the tidal tolerance parameter $\delta=10^{-5}$ and the relative and absolute accuracy of the integrator to $10^{-9}$.
We use the \fewbody default criteria to determine when to terminate each integration.

In order to detect resonant encounters, \fewbody uses the criterion of \citet{mcmillanBinarySingleStarScatteringVI1996} which counts the local minima $N_\mathrm{min}$ of the sum of the square distances between all pairs of particles.
Two minima are counted as distinct only if the local maximum in between them is at least twice that at either minimum.
Using this criterion, we categorize an integration as non-resonant if $N_\mathrm{min}=1$ and resonant if $N_\mathrm{min}>1$.
At the beginning and end of each encounter, we calculate the Milankovitch state vectors $\vect{j}$ and $\vect{e}$ and calculate $\cosam$ for each encounter from the initial and final values of $\vect{j}$.

We conduct all experiments with $\vinf=0.01\vcrit$, so that the kinetic energy of the encounter is negligible compared to the internal energy of the binary.
We also conduct all experiments in the point-particle, Newtonian limit. 
For all experiments, we draw the impact parameters from a distribution uniform in $b^2$ with the maximum value corresponding to $\rp=8\abin$.
We discuss the convergence of our results in Appendix~\ref{app:convergence}.

We fix the binary masses $m_1=m_2$ for simplicity. 
We set $\msin=km_1$, with $k$ ranging from $0.1$ to $1$ with increments of $0.1$.
For each set of masses, we conduct five sets of scattering experiments.
For four of them, we set $\ebin$ to fixed values $0.3$, $0.5$, $0.7$, and $0.9$.
For the fifth, we sample $\ebin$ from a thermal distribution.
For each ensemble, we run between $10^4$ and $10^5$ encounters.

Unless otherwise specified, all statistical fits were performed with the \texttt{curve\_fit} function from the \texttt{scipy} package \citep{scipy}.

\subsection{Comparison to Approximations}

\begin{figure*}
    \centering
    \includegraphics[width=0.92\linewidth]{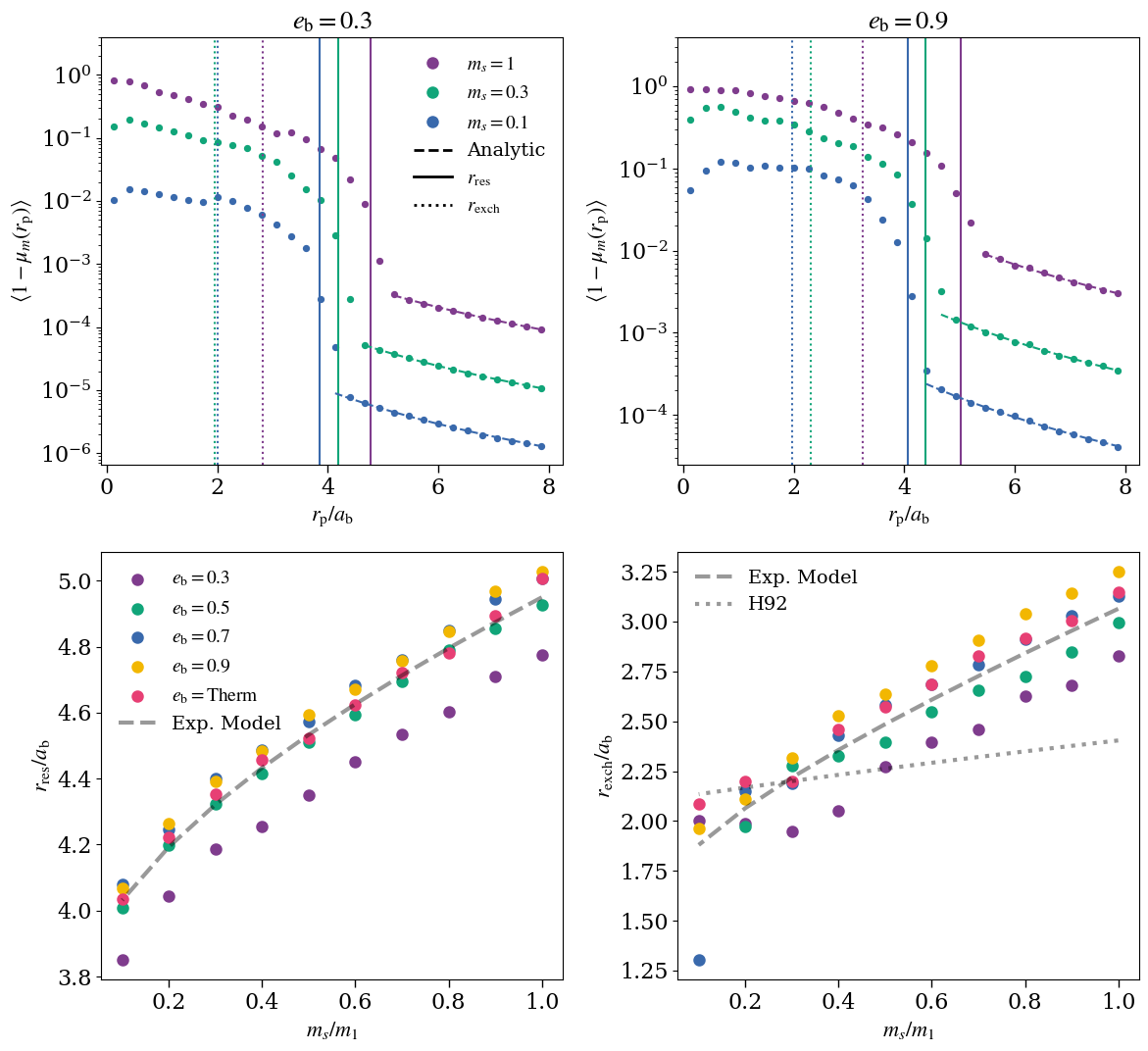}
    \caption{
    (top) The mean inclination change due to a single encounter as a function of $\rp/\abin$.
    On the y-axis, we show the binned average of $1-\mum$ for two different values of initial $\ebin$, where $\mum=\cosam$ and $\alpha_m$ is the tilt of $\lbinvec$ due to a single encounter.
    $1-\mum=1$ corresponds to an isotropic distribution of final orientations and $1-\mum\ll1$ corresponds to alignment with the initial orientation.
    In both panels, the purple, green, and blue markers and lines correspond to $\msin=1,\,0.3,\,0.1$, respectively.
    The markers show a binned average over the data.
    The dashed lines show an average calculated over the same bin widths assuming an $r^{-3}$ power-law, as implied by Eq.~\ref{eqn:weak_limit}.
    The solid and dotted vertical lines show $\rres$ and $\rexch$, representative of the maximum value of $\rp$ for resonant encounters and exchange encounters.
    The precise definitions are given in the text.
    (bottom) $\rres$ (left) and $\rexch$ (right) as a function of $\msin$.
    The different colors correspond to different values of $\ebin$. 
    Both sets of points were roughly fit with exponential models (grey dashed) ignoring the eccentricity dependence.
    The form of the model and fit parameters are described in the text.
    We also show a suggested functional form from 
    \citet{hills1991} for comparison with $\rexch$ (grey dotted).
    }
    \label{fig:rp_compare}
\end{figure*}

\begin{table*}[]
\centering
\caption{Properties of Scattering Ensembles}
\begin{tabular}{CC||CCCCC|CC|CC}
\msin/m_1 & \ebin & N_\mathrm{tot} & N_\mathrm{pres} & N_\mathrm{ex} & N_\mathrm{pres}(\rp\leq\rexch) & N_\mathrm{pres}(\rp\leq\rres) & r_\mathrm{exch} & r_\mathrm{res} & C_\mathrm{pred} & C_\mathrm{meas}  \\ \hline \hline
1.0 & 0.3 & 58977 & 51633 & 7344 & 13797 & 27945 & 2.83 & 4.78 & 0.036 & 0.045 \\
1.0 & 0.5 & 55082 & 47596 & 7486 & 13271 & 26463 & 2.99 & 4.93 & 0.11 & 0.093 \\
1.0 & 0.7 & 99998 & 84878 & 15120 & 24492 & 47485 & 3.13 & 5.01 & 0.26 & 0.27 \\
1.0 & 0.9 & 93302 & 78016 & 15286 & 22985 & 43355 & 3.25 & 5.03 & 0.72 & 1.47 \\
1.0 & \mathrm{Therm.} & 25269 & 21526 & 3743 & 6290 & 12064 & 3.15 & 5.01 & 0.39 & 1.96 \\
0.9 & 0.3 & 99998 & 89796 & 10202 & 23469 & 48764 & 2.68 & 4.71 & 0.030 & 0.037 \\
0.9 & 0.5 & 99997 & 88560 & 11437 & 24701 & 49499 & 2.85 & 4.86 & 0.093 & 0.078 \\
0.9 & 0.7 & 47332 & 41369 & 5963 & 12131 & 23480 & 3.03 & 4.94 & 0.22 & 0.24 \\
0.9 & 0.9 & 42430 & 36740 & 5690 & 11068 & 20557 & 3.14 & 4.97 & 0.60 & 1.20 \\
0.9 & \mathrm{Therm.} & 26621 & 23353 & 3268 & 6748 & 13058 & 3.00 & 4.89 & 0.32 & 2.20 \\
0.8 & 0.3 & 100000 & 91924 & 8076 & 25043 & 49548 & 2.62 & 4.60 & 0.025 & 0.031 \\
0.8 & 0.5 & 99999 & 91024 & 8975 & 25249 & 50820 & 2.73 & 4.79 & 0.076 & 0.064 \\
0.8 & 0.7 & 99999 & 90197 & 9802 & 26983 & 51061 & 2.91 & 4.85 & 0.18 & 0.19 \\
0.8 & 0.9 & 76457 & 68366 & 8091 & 21284 & 38431 & 3.04 & 4.85 & 0.49 & 1.00 \\
0.8 & \mathrm{Therm.} & 82697 & 74762 & 7935 & 22521 & 41497 & 2.92 & 4.78 & 0.26 & 1.66 \\
0.7 & 0.3 & 99997 & 94180 & 5817 & 25097 & 50794 & 2.46 & 4.53 & 0.020 & 0.024 \\
0.7 & 0.5 & 99998 & 93705 & 6293 & 26766 & 52217 & 2.65 & 4.69 & 0.061 & 0.051 \\
0.7 & 0.7 & 99998 & 92870 & 7128 & 27604 & 52238 & 2.78 & 4.76 & 0.14 & 0.15 \\
0.7 & 0.9 & 43906 & 40571 & 3335 & 12589 & 22766 & 2.91 & 4.76 & 0.39 & 0.79 \\
0.7 & \mathrm{Therm.} & 56711 & 52781 & 3930 & 16121 & 29616 & 2.83 & 4.72 & 0.21 & 1.39 \\
0.6 & 0.3 & 100000 & 96271 & 3729 & 26318 & 52029 & 2.40 & 4.45 & 0.015 & 0.019 \\
0.6 & 0.5 & 100000 & 95871 & 4129 & 27862 & 53537 & 2.55 & 4.59 & 0.046 & 0.039 \\
0.6 & 0.7 & 78054 & 74404 & 3650 & 22615 & 42123 & 2.68 & 4.68 & 0.11 & 0.11 \\
0.6 & 0.9 & 99993 & 95070 & 4923 & 29666 & 53378 & 2.78 & 4.67 & 0.30 & 0.60 \\
0.6 & \mathrm{Therm.} & 89505 & 85466 & 4039 & 25970 & 47426 & 2.69 & 4.62 & 0.16 & 1.26 \\
0.5 & 0.3 & 100000 & 97961 & 2039 & 26399 & 52548 & 2.27 & 4.35 & 0.011 & 0.013 \\
0.5 & 0.5 & 100000 & 97704 & 2296 & 27629 & 54201 & 2.39 & 4.51 & 0.033 & 0.028 \\
0.5 & 0.7 & 54925 & 53487 & 1438 & 16520 & 30063 & 2.58 & 4.57 & 0.079 & 0.082 \\
0.5 & 0.9 & 93252 & 90658 & 2594 & 28173 & 50729 & 2.64 & 4.59 & 0.21 & 0.44 \\
0.5 & \mathrm{Therm.} & 99992 & 97382 & 2610 & 29693 & 53895 & 2.57 & 4.52 & 0.12 & 0.81 \\
0.4 & 0.3 & 100000 & 99095 & 905 & 24950 & 52462 & 2.05 & 4.25 & 7.27 \times 10^{-3} & 8.85 \times 10^{-3} \\
0.4 & 0.5 & 99998 & 98930 & 1068 & 28087 & 54077 & 2.32 & 4.42 & 0.022 & 0.019 \\
0.4 & 0.7 & 99999 & 98705 & 1294 & 29093 & 54605 & 2.43 & 4.49 & 0.053 & 0.054 \\
0.4 & 0.9 & 99998 & 98535 & 1463 & 30546 & 54727 & 2.53 & 4.48 & 0.14 & 0.29 \\
0.4 & \mathrm{Therm.} & 86569 & 85504 & 1065 & 25617 & 47443 & 2.46 & 4.46 & 0.077 & 0.59 \\
0.3 & 0.3 & 99999 & 99678 & 321 & 24055 & 51813 & 1.95 & 4.19 & 4.27 \times 10^{-3} & 5.26 \times 10^{-3} \\
0.3 & 0.5 & 99998 & 99657 & 341 & 27821 & 53459 & 2.28 & 4.32 & 0.013 & 0.011 \\
0.3 & 0.7 & 61846 & 61525 & 321 & 16495 & 33468 & 2.19 & 4.40 & 0.031 & 0.031 \\
0.3 & 0.9 & 99997 & 99404 & 593 & 28332 & 54239 & 2.32 & 4.39 & 0.084 & 0.17 \\
0.3 & \mathrm{Therm.} & 59027 & 58773 & 254 & 16064 & 31931 & 2.20 & 4.35 & 0.045 & 0.39 \\
0.2 & 0.3 & 100000 & 99947 & 53 & 24521 & 50090 & 1.99 & 4.04 & 1.98 \times 10^{-3} & 2.41 \times 10^{-3} \\
0.2 & 0.5 & 100000 & 99921 & 79 & 24538 & 52343 & 1.97 & 4.20 & 6.07 \times 10^{-3} & 5.01 \times 10^{-3} \\
0.2 & 0.7 & 99999 & 99887 & 112 & 26466 & 52806 & 2.15 & 4.25 & 0.014 & 0.015 \\
0.2 & 0.9 & 99998 & 99817 & 181 & 25958 & 53000 & 2.11 & 4.26 & 0.039 & 0.078 \\
0.2 & \mathrm{Therm.} & 67463 & 67377 & 86 & 18573 & 35663 & 2.20 & 4.22 & 0.021 & 0.19 \\
0.1 & 0.3 & 100000 & 99997 & 3 & 24956 & 47963 & 2.00 & 3.85 & 5.20 \times 10^{-4} & 6.33 \times 10^{-4} \\
0.1 & 0.5 & 100000 & 100000 & 0 & 0 & 49748 & - & 4.01 & 1.59 \times 10^{-3} & 1.32 \times 10^{-3} \\
0.1 & 0.7 & 100000 & 99992 & 8 & 16239 & 50786 & 1.30 & 4.08 & 3.78 \times 10^{-3} & 3.77 \times 10^{-3} \\
0.1 & 0.9 & 100000 & 99976 & 24 & 24229 & 50663 & 1.96 & 4.07 & 0.010 & 0.020 \\
0.1 & \mathrm{Therm.} & 99997 & 99983 & 14 & 25973 & 50574 & 2.09 & 4.04 & 5.51 \times 10^{-3} & 0.038 \\
\end{tabular}
\tablecomments{Column 7: Representative maximum value of $\rp$ for exchange interactions. A precise definition is given in the text. Column 8: Same as the previous column, but for resonant interactions. Column 9: $C$ as predicted by Eq.~\ref{eqn:C}. For thermal $\ebin$, the eccentricity factor was set equal to $1$. Column 10: $C$ as measured from the data with the procedure described in the text.}
\label{tab:rp_compare}
\end{table*}

Figure~\ref{fig:rp_compare} shows the behavior of $\mum$ with respect to $\rp$. 
Table~\ref{tab:rp_compare} reports all the data generated in Fig.~\ref{fig:rp_compare}, along with relevant branching ratios.
We show two examples of the quantity $\ave{1 - \mum}$, where the average is done over fixed-width bins.
The analytic comparison (dashed) was constructed by assuming the data follows the form of Eq.~\ref{eqn:weak_limit} and estimating $C_\mathrm{meas}$ as follows.
We binned the subset of the data where $\rp/\abin>6$, computed the expected average over each bin assuming a $r^{-3}$ power law, extracted the value $C_\mathrm{meas}$ from each bin, and took the median of each estimate.
In general, $C_\mathrm{pred}$ (i.e., the value computed from Eq.~\ref{eqn:C}) was correct to within a factor of order unity, with the largest discrepancy for the $\ebin=0.9$ case (see Table~\ref{tab:rp_compare}).
We expect that the disagreement is greater when $\vinf$ is not sufficiently small compared to $\vcrit$ since there are additional terms of order the outer orbit's (hyperbolic) eccentricity.

We define $\rres$ and $\rexch$ as the 98\textsuperscript{th} percentile value of $\rp$ for encounters classified as resonant or exchanges, respectively, motivated by the inherent stochasticity in the sampling of the relative orientations of the binary and single.
However, in the latter case of exchanges, we use the second largest value of $\rp$ if the sample size is less than $50$ and the largest if the sample size is less than $20$. 
This adjustment was necessary because the exchange cross section is strongly suppressed at low $\msin/m_1$.

The distance at which Eq.~\ref{eqn:weak_limit} fails roughly corresponds to $\rres$.
At this distance, the single may be temporarily captured by the binary due to the energy exchange from single pericenter passage.
We find that the shape is roughly consistent with an exponential model when the eccentricity dependence is ignored.
The model is of the form
\begin{equation}
    r_\mathrm{model} = A_\mathrm{fit}\exp\left(B_\mathrm{fit}\sqrt{\msin/m_1}\right)
    \label{eqn:r_model}
\end{equation}
For $\rres$, $A_\mathrm{fit}=11/3$ and $B_\mathrm{fit}=3/10$, whereas for $\rexch$, $A_\mathrm{fit}=3/2$ and $B_\mathrm{fit}=5/7$.
However, based on the form of Eq.~\ref{eqn:energy}, the parameters $A$ and $B$ used to fit this model likely do not generalize to arbitrary $\vinf$, as the location at which the single can be captured depends sensitively on the chosen value of $\vinf/\vcrit$.
One would expect that energy transfer is easier for more eccentric orbits, and indeed, there is a weak dependence, though we do not speculate on the functional form.

The mass dependence of $\rexch$ is less expected.
The proposed functional form (dashed) does not match the $r\propto(1+\msin/\mbin)^{1/3}$ dependence reported by \citet{hills1991} (dotted).
In that study, all binaries were initialized with $\ebin=0.001$ and far fewer simulations were conducted.
While some eccentricity dependence is expected, it is not clear that it is strong (note that at low $\msin$ the scatter is non-monotonic with respect to $\ebin$--likely due to the small cross section at such low masses instead of a physical reason).
This may also partly be an artifact of the limited range of $\msin$ explored.

Lastly, we comment on the the parametric dependence of $\mum$ in the strong encounter regime in light of the calculations of Sec.~\ref{sec:strong_analytic}.
The overall normalization of the different curves follows expectation.
For larger $\msin$ and $\ebin$, $\lsin$ constitutes a greater proportion of the total angular momentum budget, so these curves have a maximum value closer to $1$.
However, the $\rp$ dependence does not follow the expectations from Sec.~\ref{sec:strong_analytic}.
At low $\rp$, we indeed see that $\ave{1-\mum}$ is initially increasing for both the $\msin=0.3$ and $0.1$ case at both values of $\ebin$.
We likely don't see this in the $\msin=1$ case due to the choice of bin size.
However, after the initial rise, the curve once again decreases, contrary to the expectation that the distribution become more isotropic with increasing $\lsin\propto\sqrt{\rp}$.
This is not entirely surprising, since many encounters are not ergodic \citep{traniIslesRegularitySea2024a}, especially beyond $\rexch$, where the resonances essentially correspond to a large number of small, correlated kicks on the timescale of the outer orbital period \citep{2026MNRAS.549ag944G}.
Thus, we disregard the $\rp$ dependence implied by Eq.~\ref{eqn: l_ratio} in what follows.

\begin{figure*}
    \centering
    \includegraphics[width=0.92\linewidth]{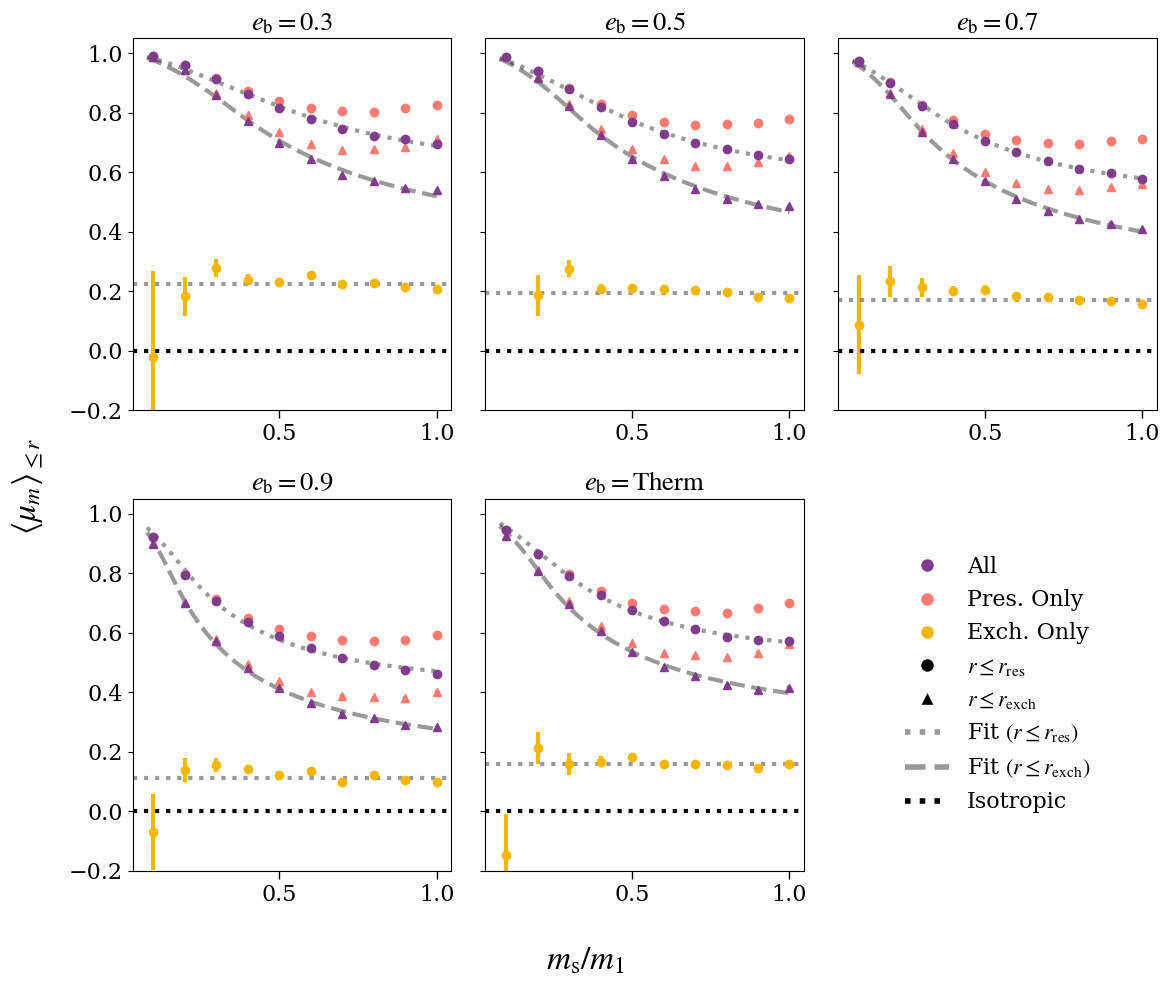}
    \caption{
    The mass dependence of the average tilt $\ave{\mum}$ for different eccentricities. 
    The purple, salmon, and yellow markers correspond to averages over both exchange and preservation interactions, only preservation interactions, and only exchange interactions, respectively. 
    The circles correspond to the averages which were restricted to $\rp\le\rres$ while the squares are similarly restricted to $\rp\le\rexch$.
    For the yellow points, we only show circles since the values are identical except where the statistics is sparse.
    We find no significant difference in $\ave{\mum}$ values when choosing the measured values of $\rres$ and $\rexch$ as the cutoff versus the exponential models presented in Fig.~\ref{fig:rp_compare}.
    We also present fits to a physically motivated model described in the text.
    }
    \label{fig:mu_mass_dependence}
\end{figure*}

\begin{table}[]
    \centering
    \caption{Fit parameters for Eq.~\ref{eqn:l_star} and straight-line fits in Fig.~\ref{fig:mu_mass_dependence}.}
    \begin{tabular}{C|CC|CC|C}
    \multicolumn{1}{c|}{ } & \multicolumn{2}{c|}{$r\leq\rres$} & \multicolumn{2}{c|}{$r\leq\rexch$} & \multicolumn{1}{c}{ }\\
    \ebin & A_\mathrm{\mu} & B_\mathrm{\mu}
           & A_\mathrm{\mu} & B_\mathrm{\mu} & C_\mathrm{\mu} \\ \hline
    0.3              & 1.55 & 2.21 & 0.61 & 1.88 & 0.22 \\
    0.5              & 1.36 & 2.32 & 0.63 & 1.78 & 0.19 \\
    0.7              & 1.34 & 2.44 & 0.62 & 1.82 & 0.17 \\
    0.9              & 1.74 & 2.64 & 0.77 & 1.76 & 0.11 \\
    \mathrm{Therm.}  & 1.85 & 2.78 & 0.76 & 2.31 & 0.16
    \end{tabular}
    \label{tab:mu_mass_dependence}
\end{table}

In Figure~\ref{fig:mu_mass_dependence}, we investigate the mass dependence of $\ave{\mum}$. 
Similar to \citet{Trani+2021}, we find that $\ave{\mum}$ is greater when conditioning on preserving the members of the original binary than when averaging over all interactions.
The difference becomes more apparent for $\msin/m_1\gtrsim0.5$, when the exchange cross section becomes non-negligible.
One might attribute this to the inclusion of many weak perturbations at large $\rp$ in the average.
However, this finding still holds for $r\leq\rexch$, though of course within this distance many of the preservation encounters will still be relatively prompt compared to the longer-lived encounters required to exchange members.
We also find that $\ave{\mum}$ for exchange-only interactions is biased away from isotropy, as expected from the phase space calculation. 
However, $\ave{\mum}$ is approximately constant for all mass ratios, in contrast with the mass-ratio dependence predicted by Eq.~\ref{eqn:mu_fixed_l}.
We calculate straight-line fits to the exchange-only data and report the resulting constant $C_\mathrm{\mu}$ in Table~\ref{tab:mu_mass_dependence}.
That eccentricity dependence of $C_\mathrm{\mu}$ is expected, since higher eccentricity increases the share of angular momentum carried by $\lbin$.
Similar findings for the eccentricity dependence were found from scattering experiment outcomes reported by \citet{samsingAGNPotentialFactories2022}.
At the lowest mass ratio, the mean inclination change appears to exceed $\pi/2$, but this is likely a sampling artifact.
The averages are still statistically consistent with both $\ave{\mum}\approx0$ and $\ave{\mum}\approx C_\mathrm{\mu}$.

Motivated by Sec.~\ref{sec:strong_analytic} and the previous discussion, we attempt to fit the total averages with models of the form of Eq.~\ref{eqn:mu_fixed_l},~\ref{eqn:mu_fixed_e}, and \ref{eqn:mu_thermal}.
However, the $\rp$ dependence of that equation clearly does not hold.
We nevertheless conjecture that an effective $\rp$ exists that would provide good agreement with the data.
In Fig.~\ref{fig:mu_mass_dependence}, the grey lines show a good agreement when Eq.~\ref{eqn:mu_fixed_l} is used in conjunction with a redefinition of the variable $l$ (Eq.~\ref{eqn: l_ratio}).
To get good agreement, we defined the new quantity
\begin{equation}
    l^\star = l\left(\ebin,\frac{r_\mathrm{model}}{A_\mathrm{\mu}\exp\left(B_\mathrm{\mu}\sqrt{\msin/m_1}\right)}\right)
    \label{eqn:l_star}
\end{equation}
where $r_\mathrm{model}$ was evaluated with the parameters reported above for both $\rres$ and $\rexch$.
For the thermal eccentricity runs, we average the eccentricity term over the thermal distribution (resulting in a factor of $2$).
$A_\mathrm{\mu}$ was allowed to vary across eccentricity in order to account for variations not captured by $r_\mathrm{model}$.
Table~\ref{tab:mu_mass_dependence} reports the best fit values.
The consistency of the fit parameters across eccentricities and both cutoff radii suggests the model is not merely overfitting.

Based on our results, we recommend the following extension of our fits for the conditioning on either preservation or exchange.
For exchanges, it seems appropriate to approximate $\ave{\mum}$ as constant. 
For preservations, we suggest using the above fit with Eq.~\ref{eqn:l_star} for $\msin/m_1<0.5$ and approximating $\msin/m_1\geq0.5$ as constant with the value given by the fit at $\msin/m_1=0.5$.
Though approximate, these fits are unlikely to produce large errors when used with Eq.~\ref{eqn:decay_time} based on the properties of the equation except near $\mu\gtrsim0.96$ where the derivative of Eq.~\ref{eqn:decay_time} is large and truncation is advisable.

\subsection{Statistical Fits}

\begin{figure*}
    \centering
    \includegraphics[width=0.92\linewidth]{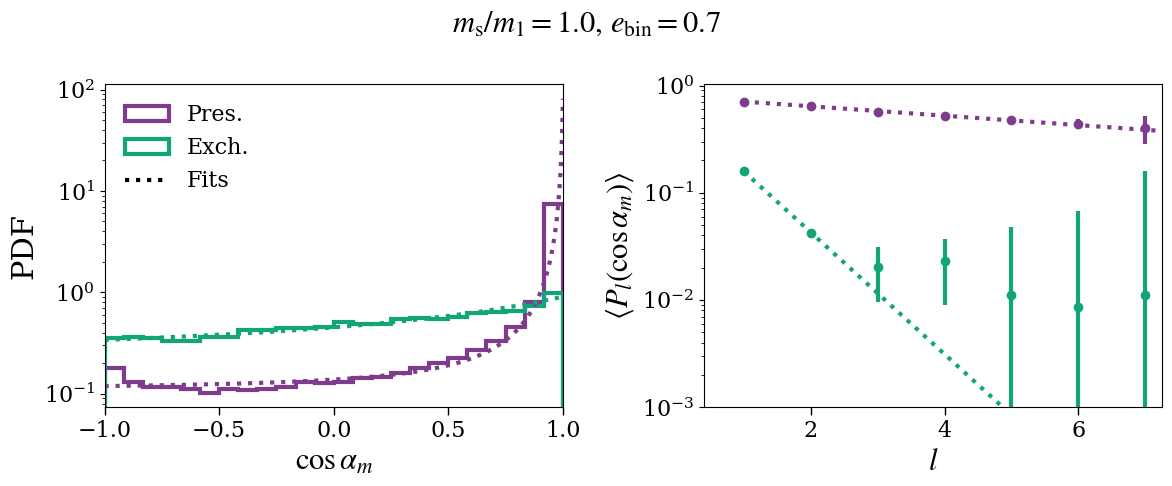}
    \includegraphics[width=0.92\linewidth]{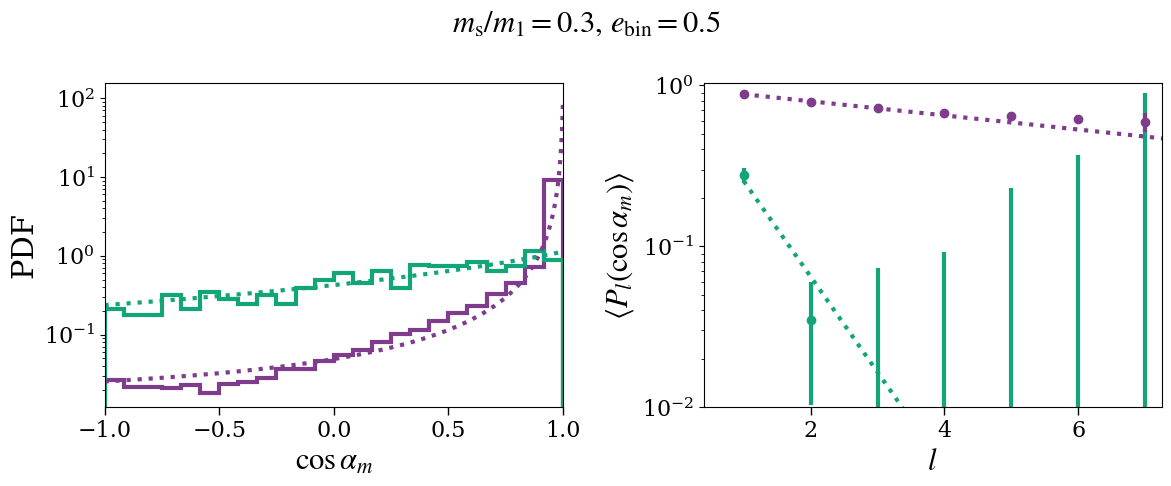}
    \includegraphics[width=0.92\linewidth]{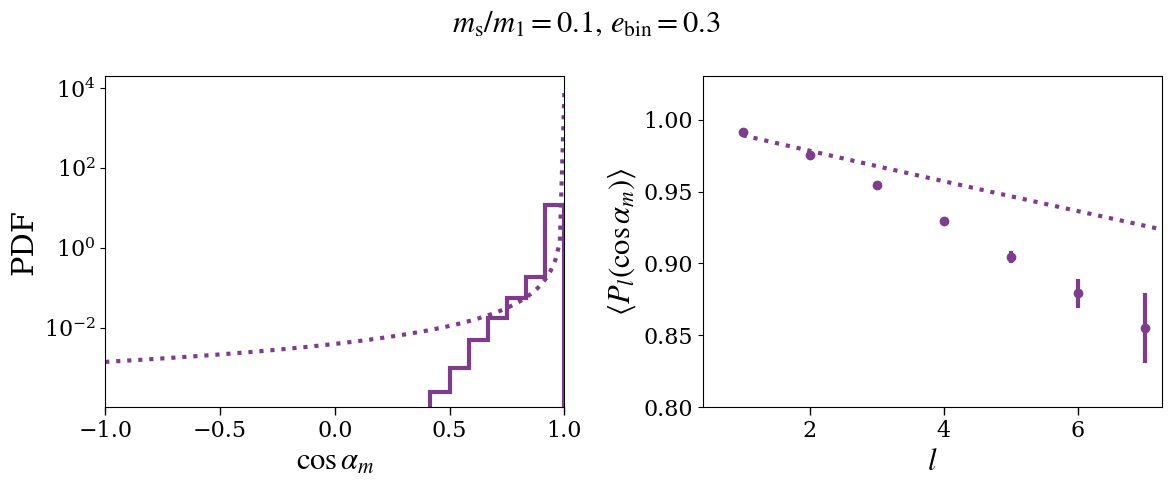}
    \caption{
    Empirical PDFs of the single encounter tilt distribution $\cosam$ within $\rres$ (left) and Legendre moments computed from distributions (right).
    The purple corresponds to the preservation distributions and the mint green to exchange distributions.
    Dotted lines show the PDF fits to the distributions obtained from fitting the Legendre moments to the analytic form.
    Each row corresponds to a different set of scattering experiments.
    In the last row, we do not include the exchange distributions because the exchange cross section is negligible.
    }
    \label{fig:dist_examples}
\end{figure*}

\begin{table}
    \caption{PDF fit parameters for $\cos\alpha_m$ distributions.}
    \label{tab:pdf_fits}
    \centering
    \begin{tabular}{CC|CC|CC|C}
    \msin/m_1 & \ebin & g_\mathrm{pres} & w_\mathrm{pres} & g_\mathrm{exch} & w_\mathrm{exch} & \mathrm{Min} \\
    \hline
    1.0 & 0.3 & 0.94 & 0.88 & 0.26 & 0.79 & -1.00 \\
    1.0 & 0.5 & 0.93 & 0.84 & 0.27 & 0.67 & -1.00 \\
    1.0 & 0.7 & 0.90 & 0.78 & 0.27 & 0.59 & -1.00 \\
    1.0 & 0.9 & 0.85 & 0.69 & 0.11 & 1.00 & -1.00 \\
    1.0 & \mathrm{Therm.} & 0.90 & 0.77 & 0.34 & 0.47 & -1.00 \\
    0.9 & 0.3 & 0.94 & 0.87 & 0.27 & 0.80 & -1.00 \\
    0.9 & 0.5 & 0.92 & 0.83 & 0.27 & 0.67 & -1.00 \\
    0.9 & 0.7 & 0.90 & 0.78 & 0.27 & 0.62 & -1.00 \\
    0.9 & 0.9 & 0.86 & 0.67 & 0.12 & 1.00 & -1.00 \\
    0.9 & \mathrm{Therm.} & 0.90 & 0.76 & 0.32 & 0.45 & -1.00 \\
    0.8 & 0.3 & 0.94 & 0.87 & 0.28 & 0.81 & -1.00 \\
    0.8 & 0.5 & 0.92 & 0.83 & 0.25 & 0.79 & -1.00 \\
    0.8 & 0.7 & 0.90 & 0.77 & 0.26 & 0.65 & -1.00 \\
    0.8 & 0.9 & 0.85 & 0.67 & 0.13 & 1.00 & -1.00 \\
    0.8 & \mathrm{Therm.} & 0.90 & 0.74 & 0.31 & 0.50 & -1.00 \\
    0.7 & 0.3 & 0.93 & 0.87 & 0.26 & 0.87 & -1.00 \\
    0.7 & 0.5 & 0.91 & 0.83 & 0.25 & 0.82 & -1.00 \\
    0.7 & 0.7 & 0.89 & 0.78 & 0.22 & 0.81 & -1.00 \\
    0.7 & 0.9 & 0.86 & 0.67 & 0.11 & 1.00 & -1.00 \\
    0.7 & \mathrm{Therm.} & 0.89 & 0.75 & 0.24 & 0.64 & -1.00 \\
    0.6 & 0.3 & 0.92 & 0.89 & 0.29 & 0.87 & -1.00 \\
    0.6 & 0.5 & 0.90 & 0.85 & 0.27 & 0.76 & -1.00 \\
    0.6 & 0.7 & 0.89 & 0.79 & 0.25 & 0.73 & -1.00 \\
    0.6 & 0.9 & 0.85 & 0.69 & 0.14 & 1.00 & -1.00 \\
    0.6 & \mathrm{Therm.} & 0.89 & 0.77 & 0.16 & 1.00 & -1.00 \\
    0.5 & 0.3 & 0.91 & 0.92 & 0.23 & 1.00 & -1.00 \\
    0.5 & 0.5 & 0.90 & 0.88 & 0.21 & 1.00 & -1.00 \\
    0.5 & 0.7 & 0.88 & 0.82 & 0.20 & 1.00 & -1.00 \\
    0.5 & 0.9 & 0.85 & 0.72 & 0.13 & 1.00 & -1.00 \\
    0.5 & \mathrm{Therm.} & 0.88 & 0.79 & 0.24 & 0.75 & -1.00 \\
    0.4 & 0.3 & 0.91 & 0.96 & 0.24 & 1.00 & -1.00 \\
    0.4 & 0.5 & 0.90 & 0.92 & 0.21 & 1.00 & -1.00 \\
    0.4 & 0.7 & 0.88 & 0.88 & 0.19 & 1.00 & -1.00 \\
    0.4 & 0.9 & 0.85 & 0.76 & 0.14 & 1.00 & -1.00 \\
    0.4 & \mathrm{Therm.} & 0.88 & 0.84 & 0.16 & 1.00 & -1.00 \\
    0.3 & 0.3 & 0.92 & 1.00 & 0.27 & 1.00 & -1.00 \\
    0.3 & 0.5 & 0.90 & 0.97 & 0.25 & 1.00 & -1.00 \\
    0.3 & 0.7 & 0.88 & 0.93 & 0.23 & 1.00 & -1.00 \\
    0.3 & 0.9 & 0.86 & 0.83 & 0.15 & 1.00 & -1.00 \\
    0.3 & \mathrm{Therm.} & 0.89 & 0.90 & 0.17 & 1.00 & -1.00 \\
    0.2 & 0.3 & 0.96 & 1.00 & 0.13 & 1.00 & -1.00 \\
    0.2 & 0.5 & 0.94 & 1.00 & 0.54 & 0.35 & -1.00 \\
    0.2 & 0.7 & 0.91 & 0.99 & 0.21 & 1.00 & -1.00 \\
    0.2 & 0.9 & 0.87 & 0.91 & 0.12 & 1.00 & -1.00 \\
    0.2 & \mathrm{Therm.} & 0.91 & 0.95 & 0.18 & 1.00 & -1.00 \\
    0.1 & 0.3 & 0.99 & 1.00 & 0.00 & 1.00 & 0.49 \\
    0.1 & 0.5 & 0.98 & 1.00 & - & - & 0.24 \\
    0.1 & 0.7 & 0.97 & 1.00 & 0.06 & 1.00 & -1.00 \\
    0.1 & 0.9 & 0.92 & 1.00 & 0.00 & 1.00 & -1.00 \\
    0.1 & \mathrm{Therm.} & 0.95 & 1.00 & 0.00 & 1.00 & -1.00 \\
    \end{tabular}
\end{table}

If the goal were simply to evaluate Eq.~\ref{eqn:decay_time}, our previous results would suffice.
However, in some contexts it may be desirable to explicitly sample the distributions of angles, so in this section we investigate fits to the distributions of $\mum$ within $\rres$.

To fit the simulated $\cosam$ distributions, we make use of the Henyey-Greenstein (HG) distribution function,
\begin{equation}
    p_\mathrm{HG}(\mu|g) = \frac{1-g^2}{2(1+g-2g\mu)^{3/2}}.
\end{equation}
The HG distribution is a natural choice here, having originally been proposed to model the angular distribution of light scattered by interstellar dust \citep{Henyey+Greenstein1941}, another manifestation of the long-standing analogy between gravitational and electromagnetic scattering.
The parameter $g=\ave{\cos\theta}\in(-1,1)$, known as the anisotropy or asymmetry factor, is the average of the distribution and controls the height of the peak at $\mu=1$ ($\mu=-1$) for positive (negative) values of $g$.
For $g=0$, the distribution is isotropic.
The pure HG distribution has the useful property that its Legendre moments are $\ave{P_l(\cos\alpha_m)} = g^l$ \citep{Roberge1983}. 
We also construct the HG+iso distribution
\begin{equation}
    p_\mathrm{HG+iso}(\mu|g,w) =w p_{HG}(\mu|g) + (1-w) p_\mathrm{iso},
    \label{eqn:PDF_hgiso}
\end{equation}
a mixture between the HG distribution with weight $w$ and the isotropic distribution $p_\mathrm{iso} = 1/2$ with weight $(1-w)$ and has Legendre moments $\ave{P_l(\cos\alpha_m)} = wg^l$.

Since the Legendre moment spectrum is the key quantity for our calculations, we fit the first $8$ empirical Legendre moments of the distributions, truncated at $\rres$, using a method of moments approach.
We use the Akaike Information Criterion to select between the two proposed models.

Table~\ref{tab:pdf_fits} reports the results of these fits and Figure~\ref{fig:dist_examples} shows some selected examples.
For exchanges, the empirically computed moments above $l=2$ are noisy since the distribution has little small-scale angular structure.
Disagreement with the model is only apparent where the moments themselves are near $0$.
However, by eye the models describe the histograms well.
A similar statement hold for the preservation distributions.
In general, the values of the first few Legendre moments are described adequately by the model.
In the middle row, the model slightly underpredicts the distribution near $\cos\alpha_m = 0.5$ and overpredicts near $\cos\alpha_m = -0.5$.
Furthermore, there is some disagreement between the empirical and model moments past $l=5$.
In the bottom row, we show the fit with the most severe disagreement.
For two cases, the fit was extremely inadequate because the domain of the models are $[-1,1]$, whereas the data was truncated at positive values.
Here, we see the phenomenon that \citet{Stone+Leigh2019} identified as angular momentum starvation, where the low-mass single is unable to carry away sufficient angular momentum to significantly tilt the binary.
The proposed models are inadequate in this regime, though the discrepancy is small within the range of the data.
We note the minimum measured value of the distributions in Tab.~\ref{tab:pdf_fits}, from which it can be seen that the truncation of the distribution only occurred in two cases. 
The fits perform best for the thermal eccentricity distributions across all masses, which are the most relevant for the calculations that follow.

\section{Discussion}
\label{sec:discussion}

We now return to the astrophysical application of these results.
While in principle these results can be applied to the dynamics of any type of binary that experience isotropic encounters with singles, we only consider the case of BBHs here.

\subsection{An Example Calculation for GW231123-like Events}

\begin{figure}
    \centering
    \includegraphics[width=0.96\linewidth]{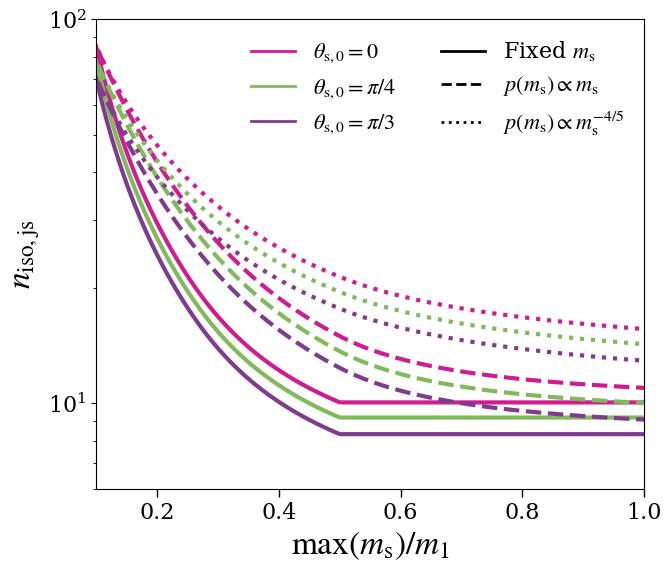}
    \caption{The number of encounters required to isotropize orientation $\nisojs$ (Eq.~\ref{eqn:decay_time_general}) as a function of the maximum $\msin$ for the distribution of singles encountered, assuming that the original binary is always preserved after each interaction. 
    The colors represent different reference vectors with pink, green, and purple corresponding to initial spin-orbit misalignment angles of $\theta_\mathrm{s,0} = 0$, $\pi/4$, and $\pi/3$, respectively.
    The solid lines correspond to fixed values of $\msin$ while dashed and dotted lines correspond to a distribution of encounters with $p(\msin)\propto\msin^\beta$ for $\beta=1$ and $\beta=-4/5$, respectively.
    }
    \label{fig:n_iso_example}
\end{figure}

\begin{figure}
    \centering
    \includegraphics[width=0.96\linewidth]{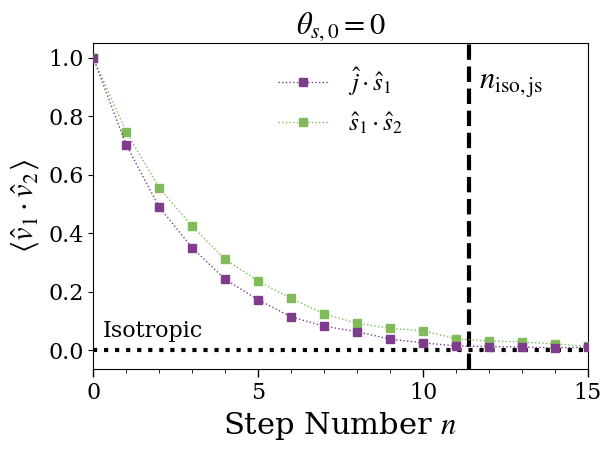}
    \includegraphics[width=0.96\linewidth]{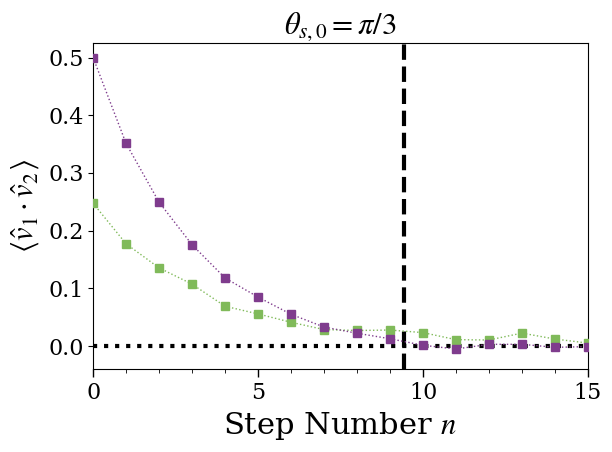}
    \caption{
    Mean cosine of spin-orbit (purple) and spin-spin (green) misalignments as a function of the random walk step number $n$ for initial spin-orbit misalignments $\theta_\mathrm{s,0} = 0$ (top) and $\pi/3$ (bottom).
    The dashed black line shows the estimate for $\nisojs$ (Eq.~\ref{eqn:decay_time_general}).
    The dotted black line shows the limit of isotropy.
    See the text for details on how the random walk trajectories were constructed.
    }
    \label{fig:walk_trajectory}
\end{figure}

\begin{figure*}
    \centering
    \includegraphics[width=0.96\linewidth]{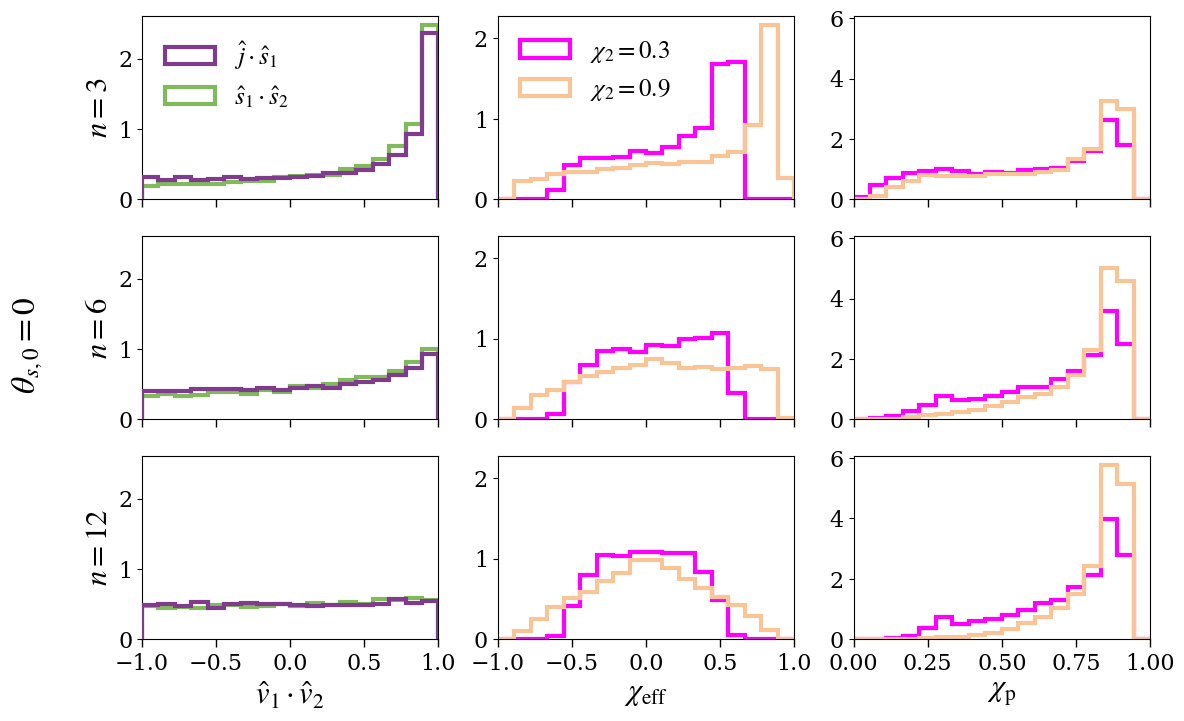}
    \includegraphics[width=0.96\linewidth]{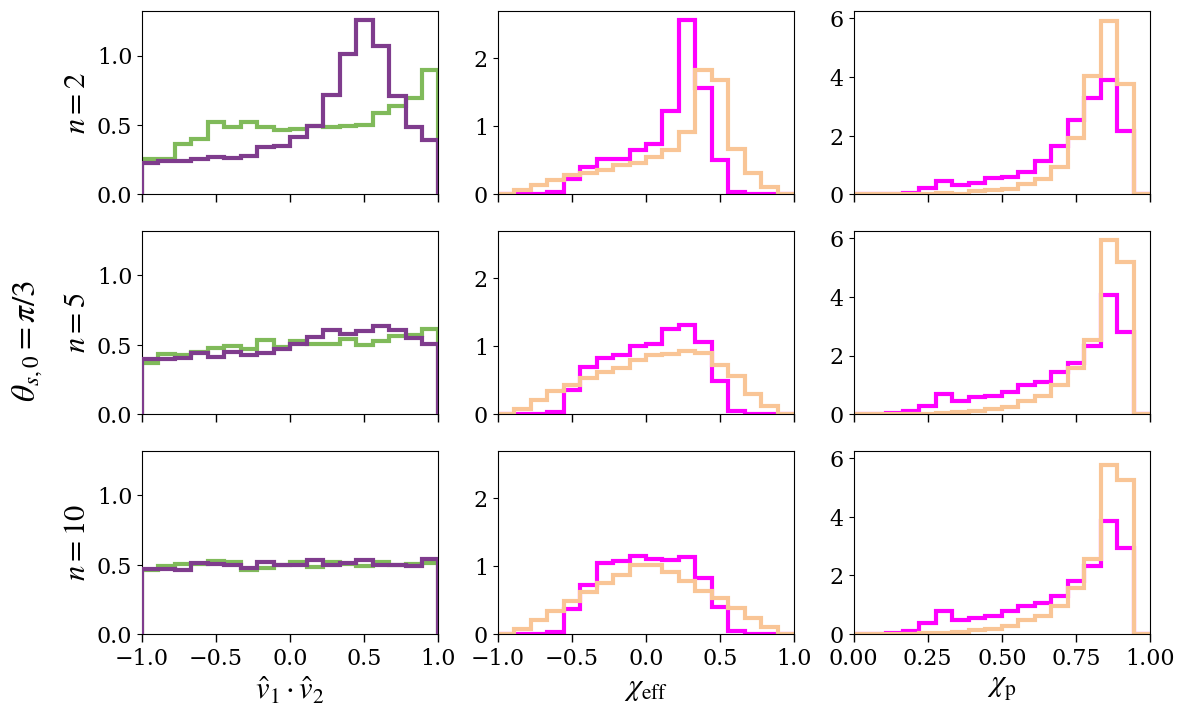}
    \caption{
    Distributions of various quantities at different step numbers (rows) of the same random walks shown in Fig.~\ref{fig:walk_trajectory}.
    The top and bottom halves correspond to initial spin-orbit misalignments $\theta_\mathrm{s,0} = 0$ and $\pi/3$, respectively.
    The chosen step numbers approximately correspond to $\nisojs/4$, $\nisojs/2$, and $\nisojs$ from top to bottom within each half.
    The left column shows the distribution of spin-orbit and spin-spin misalignments using the same colors as Fig.~\ref{fig:walk_trajectory}.
    The center and right columns show the distributions of $\chieff$ and $\chip$, respectively, for which we assume $m_1 = m_2$ and $\chi_1=0.9$.
    The magenta and apricot lines assume $\chi_2=0.9$ and $\chi_2=0.3$, respectively.
    These values roughly correspond to the properties of the progenitor BBH of the GW231123 primary suggested by \citet{Stegmann2025}.
    }
    \label{fig:walk_distributions}
\end{figure*}

As an example, we show how our work can be applied to the formation of a GW231123-like system \citep{abacGW231123BinaryBlack2025}.
The primary ($m_1\approx137\,\msun$) and secondary ($m_2\approx101\,\msun$) of GW231123 are the largest detected to date, with the primary mass confidently within or above the so-called upper mass gap due to pair instability \citep[e.g.,][]{woosley2017}.
Furthermore, both merging components have spin magnitudes $\chi_i\gtrsim0.8$.
The properties of this event pose a challenge to the canonical hierarchical merger scenario as merger products exceeding $100\,\msun$ typically involve a long chain of repeated mergers and therefore tend to spin down $\chi<0.7$ in massive clusters \citep{2026ApJ...998..138M}.
\citet{Stegmann2025} show that it is extremely difficult to form the primary of GW231123 unless the progenitor BBH of the primary had at least one highly spinning component aligned with the orbital angular momentum.
While this event may be consistent with the low-probability outcome of the canonical hierarchical merger channel \citep{passengerGW231123HierarchicalMerger2026}, it is easier to explain if at least one of the components is the remnant of a primordial BBH merger \citep{2025ApJ...994L..54P}.
A separate possibility is that BBHs with these masses and spins can be the result of spin-up due to accretion following stellar collisions \citep{kirogluBlackHoleAccretion2025, kirogluHierarchicalMergersAccretiondriven2025, kirogluSpinOrbitAlignmentMerging2025}.
In particular, \citet{kirogluSpinOrbitAlignmentMerging2025} showed that the BBH spins are generally aligned due to subsequent evolution of the accretion disks following a collision.

Now, we show how the results of Sec.~\ref{sec:theory} and \ref{sec:numerical_comparison} may be used to calculate both the number of encounters required to isotropize the spin tilt distribution and characterize the shape of the distribution prior to isotropization.
We use GW231123 for this calculation as its properties allow us to make the following simplifications.
First, the primary and secondary masses for either scenario are similar enough that our restriction of an equal-mass binary is roughly valid.
Second, due to the high masses involved, it is unlikely that there were significant numbers of BHs of comparable masses in its host environment, allowing us to put aside the possibility of exchanges. 
A more complete calculation can be done without neglecting exchanges, though in that case one must make additional assumptions about the new companion's spin orientation and magnitude.

In Figure~\ref{fig:n_iso_example}, we show $\nisojs$ evaluated at $\rres$ assuming different initial spin-orbit misalignments.
To account for the different initial spin-orbit misalignments, we used Eq.~\ref{eqn:decay_time_general} assuming a fixed misalignment angle, for which $b_1=\jhat_0\cdot\shat_i$.
For the dashed and dotted lines, we assume that the singles follow a distribution $p(\msin)\propto\msin^\beta$ following \citet{2025A&A...697A.118R}, with $\beta=1$ for a low metallicity environment and $\beta=-4/5$ for a high metallicity environment. 
Due to the mass loss from winds, high mass stars such as the likely progenitors of GW231123 are extremely unlikely to come from high metallicity environments.
We only include the $\beta=-4/5$ case here as an illustrative example.
These assumed power-law distributions do not take into account, e.g. the paucity of certain masses of black-holes, though since $\mum$ is roughly constant above $\msin/m_1 \geq 0.5$, we do not expect this to significantly impact the calculation.
We also assumed that the population of BBHs had an initially thermal $\ebin$ distribution.
Depending on the assumed formation mechanism, this assumption leads to an underprediction of $\niso$ since it is harder to torque a binary with small $\ebin$.

First, consider the accretion scenario, where the masses and spins are both the result of accretion.
The orthodox estimates of the lower edge of the BH mass gap ($\sim40\,\msun$) would imply that for this example, we should take $\msin/m_1 \sim 0.3-0.4$ as representative.
For the case where all singles lie in this mass range, we find that $\sim10$ encounters are required to isotropize $\jhat_n$, no matter the reference vector chosen.
Taking into account the BH mass spectrum, the number required increases by a factor of $\sim2-3$ for mass distributions representative of both low- and high-metallicity environments \citep{kremer+2019, yeMassDistributionBinary2026}.
Now, consider the hierarchical scenario, where at least one of the components is the result of a previous BBH merger.
\citet{Stegmann2025} suggest that a BBH with $m_1\approx m_2 \approx80\,\msun$ and at least one highly spinning component could result in a remnant with properties similar to the GW231123 primary if the spins remain aligned.
If the progenitor BBH were the result of accretion, then we could take $\msin/m_1 \sim 0.5$ as representative using the orthodox estimate of the BH mass gap.
However, if the progenitor was a primordial BBH, then that requires that the upper mass gap is much narrower than expected and $\msin/m_1 \sim 1$.
In any case, $\nisojs\approx10-20$, consistent with the accretion scenario estimate above.

To validate Eq.~\ref{eqn:decay_time_general} and characterize the full distribution 
of misalignments, we explicitly construct the random walk.
We assume $p(\msin)\propto\msin$ with $\max(\msin/m_1) = 1$ and draw $3000$ samples 
from this distribution.
For each sample, we construct the step distribution using values of $g_\mathrm{pres}$ 
and $w_\mathrm{pres}$ interpolated from the thermal-eccentricity fits in Tab.~\ref{tab:pdf_fits}, 
and draw $10000$ steps from the resulting PDF (Eq.~\ref{eqn:PDF_hgiso}).
For the random walk itself, we draw the steps for $\jhat$ from the previous 
distribution and, prior to each step, rotate $\shat_1$ and $\shat_2$ about 
$\jhat$ by a uniformly distributed random phase.

We show the outcome of $10000$ realizations of this random walk in Figure~\ref{fig:walk_trajectory} and Figure~\ref{fig:walk_distributions}.
In Fig.~\ref{fig:walk_trajectory}, we see that Eq.~\ref{eqn:decay_time_general} almost exactly predicts the decay to isotropy for both initial spin-orbit misalignments. 
We also see that the spin-spin correlations persist more than the spin-orbit correlations, contrary to the naive expectation of Eq.~\ref{eqn:decay_time_ss}.
We expect that this is a result of the large variance in the step distribution, specifically the large concentration of steps near $\alpha_m=0$.

In the top half of Fig.~\ref{fig:walk_distributions}, we see the explicit evolution from a bias towards spin-orbit alignment to isotropy.
From the shapes of the tilt distributions in the left column, we can see that the distribution of misalignments can be described by a combination of an isotropic subcomponent and a peak centered at $\cos\theta=1$, with the peak becoming less sharp and decreasing in relative weight with each encounter.
Beginning with some initial spin-orbit misalignment merely shifts the location of this peak, as shown in the bottom half of Fig.~\ref{fig:walk_distributions}.
The features of the distribution of misalignments carry over into the distributions of $\chieff$ and $\chip$.
In particular, the $\chieff$ distribution inherits the peak from the left column, with its location dependent on the assumed values for the masses and spins.
On the other hand, the $\chip$ distribution develops a peak that grows with successive encounters, reflecting the increasing concentration of misalignments near  $\sin\theta=1$ as the distribution approaches isotropy.

We showed one example of the random walk process here, but the qualitative features are general for any reasonable combination of binary and single masses, spin magnitudes, initial spin-orbit misalignments, and step distributions.
We should address one caveat for the quantitative result, namely that the \textit{number} of encounters as defined in Eq.~\ref{eqn:decay_time_general} is a somewhat ill-defined quantity since it depends on the physical cross-section, i.e. the maximum value of $\rp$, that is used to define an encounter.
We can understand the quantitative behavior of the peak as dependent on the proportion of strong and weak encounters.
However, as we know from the results of \citet{Stone+Leigh2019} and from the exchange results in Fig.~\ref{fig:mu_mass_dependence} and Fig.~\ref{fig:dist_examples}, this preference exists even for strong encounters.
Moreover, since $b^2\propto\rp$ in the parabolic limit, BBHs in dense environments are also perturbed by weak encounters prior to merger.
As we show in Fig.~\ref{fig:rp_compare}, our choice of $\max(\rp)=\rres$ is a reasonable choice to capture the impact of both strong and weak encounters.
Furthermore, as we show in App.~\ref{app:convergence}, this choice only leads to a slight overestimate of the rate of isotropization.

\subsection{Preservation of Spin-Orbit Alignment in the Broader BBH Population}

We now turn to the broader implications.
While the above calculation only examined the spin tilt evolution of a single BBH, the conclusion generalizes to a population: an aligned BBH may retain its alignment if it experiences a sufficiently small number of encounters.
This requires that for at least some initially aligned BBHs, $\abin$ is initially small.

This is likely the case for primordial BBHs.
If a primordial BBH is born from chemically homogeneous evolution, then the two components will be highly spinning with an extremely tight initial separation \citep{mandelMergingBinaryBlack2016, deminkChemicallyHomogeneousEvolutionary2016, marchantNewRouteMerging2016}.
Similarly, spin-up due to mass transfer or tidal spin up prior to the second supernova is generally only efficient at small separations \citep[e.g.,][]{baveraImpactMasstransferPhysics2021}.
On the other hand, it is thought that in high mass stars, angular momentum is efficiently transported from the core to the envelope, resulting in a low natal $\chi$ in the absence of the aforementioned spin-up mechanisms \citep{fuller+ma2019}.
Thus, one would expect that primordial BBHs with non-negligible spins should be both initially aligned and relatively compact, experiencing few strong encounters prior to merger.

This is supported by recent simulations by \citet{oconnorBlackHoleMergers2026} using the Monte Carlo code \texttt{CMC}.
In less compact clusters, many primordial BBHs remained intact until merger while experiencing few to no strong gravitational interactions.
The primordial BBH mergers were also strongly biased towards the lower masses, which were less dynamically active compared to their heavier counterparts.
Their simulations also demonstrate that if primordial BBHs in dense environments are a non-negligible contribution to the alignment signature reported by the LVK Collaboration, this signature should have a unique redshift evolution compared to other features in the population.

A similar logic holds for accretion-induced alignment.
\citet{kirogluSpinOrbitAlignmentMerging2025} show that BBH-star collisions can occur for a wide range of BBH separations $\sim 0.02 - 10\,\mathrm{au}$.
However, the spins will only be aligned if the BBH separation is sufficiently small $\lesssim 1\,\mathrm{au}$ at the time of the collision since close pericenter passages are required to torque the accretion disks.
Moreover, over the course of the interaction, $\abin$ can shrink by up to a factor of $10$, reducing the probability of further encounters.

\subsection{Caveats and Future Work}
\label{sec:caveats}

We highlight several assumptions we made here.
The most pressing future extension is to generalize our numerical results beyond the equal-mass binary assumption.
In general, BBHs from any channel can have a wide range of mass ratios.
Furthermore, many of the fits reported here are not necessarily valid in the limit where $m_1\gtrsim m_2\gtrsim\msin$, such as could be the case when considering intermediate-mass BH interactions in dense environments \citep{2026arXiv260223431M}.

A second open question concerns the role of binary--binary encounters.
It is common to assume that the rate of BH binary--binary encounters in dense clusters is suppressed since the typical outcome of such encounters is to destroy one of the initial binaries \citep{2016MNRAS.456.4219A,leighChaoticFourbodyProblem2016,marinpinaDemographicsThreebodyBinary2024, barreraretamalChaoticFourbodyProblem2024, 2025A&A...698A.229M}.
However, if the binary fraction of BH progenitors is large, as implied by recent observations \citep{offnerOriginEvolutionMultiple2023}, then at least some fraction of the initial population of primordial BBHs should be expected to interact with each other soon after BH formation.
Expectations for orientation change due to binary--binary encounters are unclear.
In the case of weak interactions, \citet{hamers+samsing2020} suggest that Eq.~\ref{eqn:weak_limit} is still valid for sufficiently large $\rp$.
The case of strong interactions is much less clear.
Motivated by the heuristic that the $2+1+1$ (single binary ionization) outcome can be modeled as the successive ejection of the two ionized binary members \citep[e.g.,][]{leighChaoticFourbodyProblem2016}, \citet{Trani+2021} speculated that the outcome of a binary--binary interaction would be similar to two successive binary--single interactions.
While a similar prograde preference may be present for binary--binary encounters, 
it is likely much weaker than in the binary--single case, since the greater range 
of allowed configurations dilutes the phase space constraint that drives alignment in 
the latter.

One of the key assumptions underlying our study was the implicit assumption that the binary--single encounters, and therefore the steps of the random walk, are isotropic.
However, it is known that many clusters have observational signatures consistent with bulk rotation \citep[e.g.,][]{petraliaSignatureSystemicRotation2024}.
As a result, a number of studies have studied the impact of initial rotation on the evolution of clusters \citep[e.g.,][]{kamlahImpactStellarEvolution2022, bianchiniROLLINRotatingGlobular2026}.
While it is expected that the high-mass components (i.e. the BHs and their progenitors) should quickly reach an isotropic configuration \citep{kamlahImpactStellarEvolution2022}, it is possible that some fraction of the merging population of BBHs experienced their dynamical hardening phase during an initial period when there was a preferred orientation for encounters.
Furthermore, in AGN disks, the encounter orientations are expected to be extremely biased within the plane of the disk.
\citet{samsingAGNPotentialFactories2022} simulated a small set of binary--single encounters with fixed inclinations for a single set of masses.
Generalizing the random walk model of Sec.~\ref{sec:randomwalk} would require finding the discrete equivalent of the anisotropic diffusion tensor, which itself would require more scattering experiments in the spirit of \citet{samsingAGNPotentialFactories2022}.

The final limitation we highlight is the neglect of finite-size effects and short-range forces.
Including these effects could bias the inclination-change distributions and alter our fits.
For the current study, we focused on the case of BBHs.
The main effect of explicitly including relativistic terms in the equations of motion is to introduce the possibility of mergers during the scattering encounters.
However, such mergers, referred to as captures, are typically the result of the formation of extremely eccentric intermediate binaries during resonant encounters \citep[e.g.,][]{samsingFormationEccentricCompact2014}, which are relatively rare compared to the probability of a resonant encounter happening at all.
As shown in previous work by \citet{2026arXiv260223431M}, the inclusion of relativistic effects only significantly changed the distribution of post-encounter energy changes well past the limit by which typical BBHs would have merged due to successive encounters.
We expect the same to hold for the inclination kicks.
However, it is possible that during such captures, the spin terms interact with the orbital angular momentum in a manner similar to hierarchical triples \citep[e.g.,][]{antonini+2018}.

\section{Summary and Conclusion}
\label{sec:summary}

In this work, we studied the evolution of the spin-orbit misalignment of BBHs through binary--single encounters in dense stellar systems.
The key results are as follows:

\begin{enumerate}
    \item We modeled the evolution of the BBH orbital angular momentum orientation through successive binary--single encounters as an isotropic random walk on the unit sphere.
    The evolution of the distribution of orientations depends only on the Legendre series coefficients of the step distribution.
    From this model, we derived a closed-form expression for the number of encounters required to reach isotropy, depending only on the mean inclination change per encounter.
    \item We conducted a large suite of Newtonian point-particle binary--single scattering experiments with equal-mass binaries, with the mass of the encountering single and the eccentricity of the binary varied across ensembles.
    We confirmed the preference, first noted by \citet{Stone+Leigh2019} and \citet{Trani+2021}, for preserving the initial orientation of the binary orbital angular momentum \textit{even when only considering strong interactions}.
    While this preference is stronger for interactions that preserve the members of the initial binary, it is still present for interactions resulting in an exchange.
    We presented semianalytic fits to both the mean inclination change as a function of mass ratio and the full PDF of inclination changes, enabling both analytic estimates and direct Monte Carlo sampling.
    \item We applied these results to the astrophysical population of merging BBHs. 
    We presented an example calculation for potential progenitors of GW231123, though the qualitative findings hold more generally.
    Given initially aligned spins, the degree of surviving spin-orbit alignment depends on the number of encounters prior to merger.
    We found that, even in environments with no preferred orientation, initial spin-orbit alignment can survive several strong encounters before being erased.
    In general, understanding the decay of alignment in isotropic environments is key to interpreting current and future observations by the LVK Collaboration.
\end{enumerate}

\begin{acknowledgements}
    The authors would like to thank Anarya Ray, Nicholas Stone, Michael Zevin, Elena Gonzalez-Prieto, Charles Gibson, and Yiqiu Zhou for useful discussions.
    This work was supported by NSF Grants AST-2108624 and AST-2511543 at CIERA, and used computing resources funded by NSF Grants PHY-1726951 and PHY-2406802. 
    CEO and FK acknowledge support from a CIERA Postdoctoral Fellowship.
    Our research was supported in part through the computational resources and staff contributions provided for the Quest high performance computing facility at Northwestern University, which is jointly supported by the Office of the Provost, the Office for Research, and Northwestern University Information Technology.
\end{acknowledgements}

\software{
    \texttt{Fewbody} \citep{fregeauStellarCollisionsBinarybinary2004}, 
    \texttt{NumPy} \citep{numpy},
    \texttt{SciPy} \citep{scipy},
    \texttt{pandas} \citep{pandas2010, pandas2020},
    \texttt{Matplotlib} \citep{matplotlib}
}

\bibliography{refs}{}
\bibliographystyle{aasjournal}

\appendix

\section{Extension to Arbitrary Reference Vectors}
\label{app:general_walk_reference}

Suppose that instead of the initial direction $\jhat_0$, we cared about some other reference vector $\shat_i$ with a distribution $p(\nu_s,\phi_s)$ where $\nu_s=\cos\theta_s$ is the cosine of the angle between $\shat_i$ and $\jhat_0$ and $\phi_s$ is the azimuthal angle.
Now, the distribution of the dot product $\jhat_n\cdot\shat_i = \cos\theta_i$ can be written as a Legendre series expansion
\begin{equation*}
    p(\cos\theta_i) = \sum_{l=0}^\infty \frac{2l+1}{2} c_l P_l(\cos\theta_i)
\end{equation*}
with coefficients
\begin{equation*}
    c_l = \frac{4\pi}{2l+1} \sum_{m=-l}^l a_{lm} b_{lm}
\end{equation*}
by the spherical harmonics addition theorem.
Here, $a_{lm}$ and $b_{lm}$ are the spherical harmonic expansion coefficients of the distributions $p(\mu_{j,n})$ (Eq.~\ref{eqn:random_walk}) and $p(\nu_s,\phi_s)$, respectively.
However, since $p(\mu_{j,n})$ is azimuthally symmetric, only the $m=0$ terms survive,
\begin{equation*}
    c_l = \frac{4\pi}{2l+1} a_{l0} b_{l0}
\end{equation*}
Moreover, we already know from Eq.~\ref{eqn:random_walk} that
\begin{equation*}
    a_{l0} = \sqrt{\frac{2l+1}{4\pi}} \ave{P_l(\mum)}^n,
\end{equation*}
though the prefactors depend on the normalization convention.
In any case, if we also impose that $p(\nu_s,\phi_s)=p(\nu_s)$ (which is typically the case in our scenario due to the spin-orbit precession), then we find that 
\begin{equation*}
    c_l = b_l \ave{P_l(\mum)}^n
\end{equation*}
so that
\begin{equation}
    p(\cos\theta_i) = \sum_{l=0}^\infty \frac{2l+1}{2} b_l \ave{P_l(\mum)}^n P_l(\cos\theta_i), 
    \label{eqn:random_walk_general_app}
\end{equation}
analogous to Eq.~\ref{eqn:random_walk}, where
\begin{equation*}
    b_l = \int_{-1}^{1} p(\nu_s) P_l(\nu_s)\dd{\nu_s}
\end{equation*}

Now, after some number of encounters, the magnitude of one of the terms in the expansion may decay to a value $\epsilon$, i.e.
\begin{equation*}
    \frac{2l+1}{2} b_l \ave{P_l(\mum)}^n = \epsilon.
\end{equation*}
Solving for $n$, we find
\begin{equation*}
    n = \frac{\ln\left( \frac{2\epsilon}{2l+1} \right) - \ln|b_l|}{\ln \ave{P_l(\mum)}}.
\end{equation*}
Eq.~\ref{eqn:decay_time} may be recovered with the substitutions $l=1$, $\epsilon\approx0.027$, and $p(\nu_s)=\delta(\nu_s-1)$, for which $b_1 = 1$.
Thus, the analog of Eq.~\ref{eqn:decay_time} is simply
\begin{equation}
    n = -\frac{4 + \ln|b_1|}{\ln \ave{\mum}}.
    \label{eqn:decay_time_general_app}
\end{equation}

\section{Derivation of the Quadrupole Order Result}
\label{app:quadrupole}

We calculate the perturbation of the binary by the passing single during the entire encounter, 
averaged over the orbit of the binary, at the quadrupole order of approximation. 
We use the vectorial formalism for secular dynamics \citep[e.g.,][]{tremaineWhyEarthSatellites2014}, 
where the orbit of the binary is described by the eccentricity (Laplace--Runge--Lenz) and normalized angular momentum vectors:
\begin{align}
    \vect{e} &= \ebin \left[ \zhat \sin\omega \sin{I} + \xhat \left( \cos{\Omega} \cos{\omega} - \sin{\Omega} \sin{\omega} \cos{I} \right) + \yhat \left( \sin{\Omega} \cos{\omega} + \cos{\Omega} \sin{\omega} \cos{I} \right) \right], \\
    \vect{j} &= \left( 1 - \ebin^{2} \right)^{1/2} \left[ \zhat \cos{I} + \sin{I} \left( \xhat \sin{\Omega} - \yhat \cos{\Omega} \right) \right],
\end{align}
where the unit vectors $(\xhat, \yhat, \zhat)$ form a right-handed triad. 
We are construct our coordinate system 
with the origin at the three-body system's center of mass, 
$\xhat$ pointing from the origin to the third body's periapsis, 
and $\zhat$ normal to the plane of the single's orbit. 
The disturbing function then has the form \citep[e.g.,][]{liu+lai2018}
\begin{equation}
    \Phi = \frac{G \mubin \msin \abin^{2}}{4 r_{\rm out}^{3}} \left[ -1 + 6 \ebin^{2} + 3(\vect{j} \cdot \rhat_{\rm out})^{2} - 15 \left( \vect{e} \cdot \rhat_{\rm out} \right)^{2} \right],
\end{equation}
where $\vect{r}_{\rm out} = r_{\rm out} \rhat_{\rm out}$ is the position of the single relative to the center of mass. 
The resulting equations of motion are
\begin{align}
    \frac{\dd \vect{j}}{\dd t} &= \frac{3}{2} \frac{\msin}{\mbin} \left(  \frac{\abin}{r_{\rm out}} \right)^{3} n \left[ (\vect{j} \cdot \rhat_{\rm out}) (\rhat_{\rm out} \times \vect{j}) - 5 (\vect{e} \cdot \rhat_{\rm out} ) (\rhat_{\rm out} \times \vect{e}) \right], \\
    \frac{\dd \vect{e}}{\dd t} &= \frac{3}{2} \frac{\msin}{\mbin} \left(  \frac{\abin}{r_{\rm out}} \right)^{3} n \left[ (\vect{j} \cdot \rhat_{\rm out}) (\rhat_{\rm out} \times \vect{e}) - 5 (\vect{e} \cdot \rhat_{\rm out} ) (\rhat_{\rm out} \times \vect{j}) - 2 \vect{j} \times \vect{e} \right].
\end{align}
We express $\vect{r}_{\rm out}$ in a convenient form for unbound trajectories, namely
\begin{align}
    r_{\rm out}(\psi) = \frac{r_{\rm p}(1+e_{\rm out})}{1 + e_{\rm out} \cos{\psi}}, \hspace{0.5cm} \rhat_{\rm out}(\psi) = \xhat \cos{\psi} + \yhat \sin{\psi} 
\end{align}
where $\psi$ is the third body's true anomaly. 
We now integrate the equations of motion over a full unbound encounter, 
using the following differential relation between true anomaly and time:
\begin{equation}
    \frac{\dd\psi}{(1 + e_{\rm out} \cos{\psi})^{2}} = \left[ \frac{G M}{r_{\rm p}^{3} (1+e_{\rm out})^{3}} \right]^{1/2} \dd t \equiv \frac{\dd t}{T_{\rm p}}.
\end{equation}
Letting $T$ be the time elapsed between the periapsis ($\psi = 0$) and a position with true anomaly $\psi_{T}$ on the outbound portion of the orbit, 
we have
\begin{equation}
    \Delta \vect{j} = \lim_{T \to \infty} \int_{-T}^{T} \frac{\dd \vect{j}}{\dd t} \, \dd t = \int_{-\psi_{\infty}}^{\psi_{\infty}} \left( \frac{\dd \vect{j}}{\dd t} \right)_{\psi} \frac{T_{\rm p} \, \dd \psi}{(1 + e_{\rm out} \cos{\psi})^{2}},
\end{equation}
and similarly for $\Delta \vect{e}$. 
Here, $\psi_{\infty} = \arccos(-1/e_{\rm out})$ is the true anomaly at outbound infinity. 
The result depends on the following integrals: 
\begin{align}
    I_{0} &= \lim_{T \to \infty} \int_{-T}^{T} \frac{\dd t}{r_{\rm out}^{3}} = \frac{2 T_{\rm p}}{r_{\rm p}^{3} (1 + e_{\rm out})^{3}} \left( \psi_{\infty} + e_{\rm out} \sin{\psi_{\infty}} \right), \\
    I_{1} &= \lim_{T \to \infty} \int_{-T}^{T} \frac{\cos^{2}(\psi)}{r_{\rm out}^{3}} \, \dd t = \frac{T_{\rm p}}{r_{\rm p}^{3} (1 + e_{\rm out})^{3}} \left( \psi_{\infty} + \frac{\sin{2 \psi_{\infty}}}{2} + 2 e_{\rm out} \left[ \sin{\psi_{\infty}} - \frac{1}{3} \sin^{3}(\psi_{\infty}) \right] \right), \\
    I_{2} &= \lim_{T \to \infty} \int_{-T}^{T} \frac{\sin^{2}(\psi)}{r_{\rm out}^{3}} \, \dd t = \frac{1}{r_{\rm p}^{3} (1 + e_{\rm out})^{3}} \left( \psi_{\infty} - \frac{\sin{2 \psi_{\infty}}}{2} + \frac{2 e_{\rm out} \sin^{3}(\psi_{\infty})}{3} \right), \\
    I_{3} &= \lim_{T \to \infty} \int_{-T}^{T} \frac{\cos{\psi} \sin{\psi}}{r_{\rm out}^{3}} \, \dd t = 0.
\end{align}
We now take the limit $e_{\rm out} \to 1$ ($\psi_{\infty} \to \pi$), so that
\begin{equation}
    I_{0} = \frac{\pi T_{\rm p}}{4 r_{\rm p}^{3}}, \hspace{0.25cm} I_{1} = I_{2} = \frac{\pi T_{\rm p}}{8 r_{\rm p}^{3}}.
\end{equation}

In the end, the result is
\begin{align}
    \Delta \vect{j} &= \frac{3}{2} \frac{\msin}{\mbin} \left( \frac{\abin}{r_{\rm p}} \right)^{3} n T_{\rm p} \left\{ \frac{5 \pi \ebin^{2}}{8} \sin{I} \sin{\omega} \left[ \xhat \left( \sin{\Omega} \cos{\omega} + \cos{\Omega} \sin{\omega} \cos{I} \right) + \yhat \left( \cos{\Omega} \cos{\omega} - \sin{\Omega} \sin{\omega} \cos{I} \right) \right] \right. \nonumber \\
    & \hspace{5.5cm} \left. - \frac{\pi}{8} (1 - \ebin^{2}) \sin{I} \cos{I} \left(\xhat \cos{\Omega} + \yhat \sin{\Omega} \right) \right\}, \\
    &= \frac{3 \pi}{16} \frac{\msin}{\mbin} \left( \frac{\abin}{r_{\rm p}} \right)^{3} n T_{\rm p} \left[ 5 ( \vect{e} \cdot \zhat ) ( \zhat \times \vect{e} ) - (\vect{j} \cdot \zhat) ( \zhat \times \vect{j} ) \right], \\ 
    &= \frac{3 \pi}{8} \frac{\msin}{\mbin} \left( \frac{2 \mbin}{\mbin + \msin} \right)^{1/2} \left( \frac{\abin}{r_{\rm p}} \right)^{3/2} \left[ 5 ( \vect{e} \cdot \zhat) ( \zhat \times \vect{e} ) - (\vect{j} \cdot \zhat) (\zhat \times \vect{j}) \right].
\end{align} 
Similarly, for the eccentricity vector we have 
\begin{equation}
    \Delta \vect{e} = \frac{3 \pi}{8} \frac{\msin}{\mbin} \left( \frac{2 \mbin}{\mbin + \msin} \right)^{1/2} \left( \frac{\abin}{r_{\rm p}} \right)^{3/2} \left[ 5 (\vect{e} \cdot \zhat) (\zhat \times \vect{j}) - (\vect{j} \cdot \zhat) (\zhat \times \vect{e}) + 2 \vect{j} \times \vect{e} \right].
\end{equation}
These expressions can be generalized for arbitrary orientations of the orbit via the substitution $\zhat \to \nhat$, 
where $\nhat$ is the unit vector along the outer body's angular momentum axis. 
It can be shown that the scalar eccentricity and the inclination angle change correspondingly by
\begin{align}
    \Delta \ebin &= - \frac{15 \pi }{16} \frac{\msin}{\mbin} \left( \frac{2 \mbin}{\mtot} \right)^{1/2} \left( \frac{\abin}{r_{\rm p}} \right)^{3/2} \ebin (1 - \ebin^{2})^{1/2} \sin^{2}(I) \sin(2 \omega), \\ 
    \Delta I &= \frac{15 \pi }{16} \frac{\msin}{\mbin} \left( \frac{2 \mbin}{\mtot} \right)^{1/2} \left( \frac{\abin}{r_{\rm p}} \right)^{3/2} \frac{\ebin^{2}}{(1 - \ebin^{2})^{1/2}} \sin(2I)\sin(2 \omega).
\end{align}
Note $\Delta \ebin$ and $\Delta I$ vanish when $\ebin \to 0$. 
In this limit, the lowest nontrivial contribution appears at the octupole order \citep[e.g., ][]{heggieEffectEncountersEccentricity1996}.

\section{A Note on Convergence}
\label{app:convergence}

We must verify the convergence of the isotropization rate.
As with the energy hardening rate, we must verify that the isotropization rate,
\begin{equation}
    D_\mathrm{iso} = \frac{-\ln{\ave{\mum}}}{\tenc} \propto -r_\mathrm{p,max}\ln{\ave{\mum}},
    \label{eqn:diff_rate}
\end{equation}
converges as $r_\mathrm{p,max}\rightarrow\infty$,
where the final expression was computed assuming the parabolic limit.
The exact expression reduces to the diffusion rate in the small-angle limit.

We can evaluate the asymptotic value of this expression as follows.
First, recall that in the parabolic limit 
\begin{equation*}
    \ave{\mum}(r) = \frac{1}{r} \int_0^r f(r')\dd r'
\end{equation*}
From Fig.~\ref{fig:rp_compare}, we can split up the integral into a numerical and analytic (Eq.~\ref{eqn:weak_limit}) component:
\begin{equation*}
    \ave{\mum}(r) = \frac{1}{r}\left[\mu_\mathrm{meas}r_\mathrm{ref} + \int_{r_\mathrm{ref}}^r \left(1 - Cr'^{-3}\right)\,dr'\right].
\end{equation*}
Here, we define $\mu_\mathrm{meas}=\ave{\mum}(r_\mathrm{ref})$.
Note that we have assumed $r$ is in units of $\abin$ for simplicity.
Upon evaluating the integrals, the expression becomes
\begin{equation*}
    \ave{\mum}(r) = 1 - \frac{K}{r} + \frac{C}{2r^{3}}, \qquad K = r_\mathrm{ref}(1 - \mu_\mathrm{meas}) + \frac{C}{2r_\mathrm{ref}^2}.
\end{equation*}
We substitute this into Eq.~\ref{eqn:diff_rate} and use the Taylor expansion of $\ln(1+x)$ to find
\begin{equation*}
    -r\ln\langle\mu_\mathrm{m}\rangle(r) = K + \frac{K^2}{2r} - \frac{C}{2r^2} + O(r^{-3}).
\end{equation*}
Upon taking the limit, this becomes
\begin{equation}
    -\lim_{r\rightarrow\infty} r\ln{\ave{\mum}} = r_\mathrm{ref}(1 - \mu_\mathrm{meas}) + \frac{C}{2r_\mathrm{ref}^2}
    \label{eqn:diff_rate_limit}
\end{equation}
Since the quadrupole order expression is only valid for $r_\mathrm{ref}>\rres$, it is clear that, in general, the first term dominates.
In other words, the very weak interactions only contribute a small correction to the total diffusion rate.

\begin{figure}[h]
    \centering
    \includegraphics[width=0.48\textwidth]{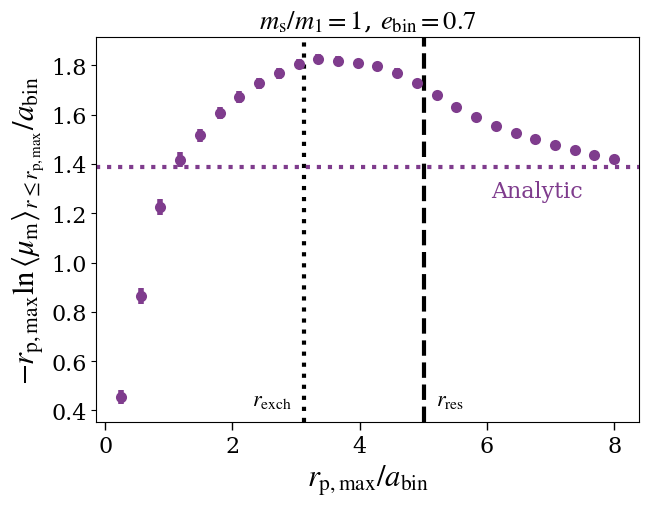}
    \hfill
    \includegraphics[width=0.48\textwidth]{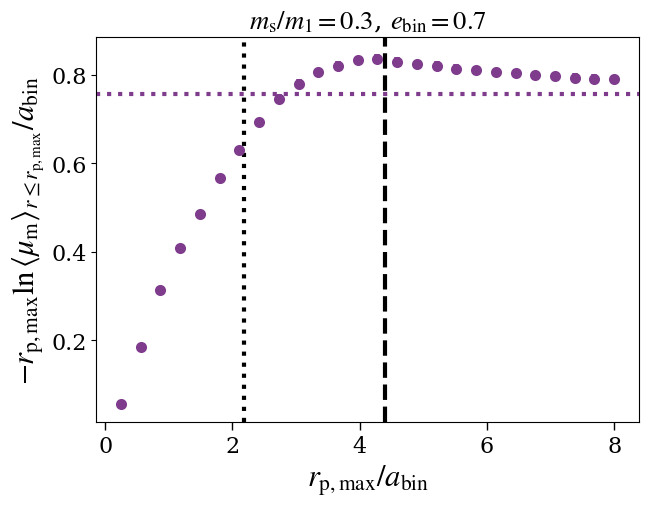}
    \caption{
    Measured isotropization rate (Eq.~\ref{eqn:diff_rate}) as a function of $r_\mathrm{p,max}$.
    For each data point (purple circles), the average was computed over the preservation interactions up to the given value of $r_\mathrm{p,max}$.
    The analytic estimate (purple dashed) was computed using Eq.~\ref{eqn:diff_rate_limit} with $r_\mathrm{ref}=6\abin$ and $C=C_\mathrm{meas}$ (see Tab.~\ref{tab:rp_compare}).
    For comparison, we also show the measured values of $\rexch$ and $\rres$ with the black dotted and dashed lines, respectively.
    }
    \label{fig:convergence}
\end{figure}

We show two examples of the convergence of this quantity in Figure~\ref{fig:convergence}.
We show these estimates only including the preservation interactions since the exchange cross section converges for sufficiently large $\rp$.
In either case, the analytic estimate seems to be adequate, though the rate of convergence is slower for lower values of $\msin$.
In Sec.~\ref{sec:discussion}, we decided to define an encounter with $r_\mathrm{p,max}=\rres$.
As we can see here, this leads to a roughly $20\%$ overestimate of the diffusion rate in the equal-mass case, though the overestimate becomes smaller for lower $\msin$.

\end{document}